\documentclass[twocolumn,showpacs,preprintnumbers,amsmath,amssymb,
               superscriptaddress]{revtex4}

\usepackage{graphicx}

\newcommand{\ba}{\begin{array}}
\newcommand{\ea}{\end{array}}

\newcommand{\req}[1]{Eq.~(\ref{#1})}

\newcommand{\dif}{{\rm d}}

\newcommand{\Dslash}{\relax{\kern+.25em / \kern-.70em D}}

\newcommand{\Real}{\relax{\mathsf{\Gamma\kern-.35em R}}}
\newcommand{\Int}{\relax{\mathsf{Z\kern-.40em Z}}}

\newcommand{\NF}{N_\mathrm{\scriptstyle f}}

\newcommand{\SF}{{\rm SF}}
\newcommand{\GF}{{\rm GF}}

\newcommand{\gbar}{\kern1pt\overline{\kern-1pt g\kern-0pt}\kern1pt}
\newcommand{\mbar}{\kern2pt\overline{\kern-1pt m\kern-1pt}\kern1pt}
\newcommand{\obar}[1]{\kern3pt\overline{\kern-2pt #1\kern-0pt}\kern1pt}

\newcommand{\lQCD}{\Lambda_{\rm\scriptscriptstyle QCD}}
\newcommand{\MW}{M_{\rm\scriptscriptstyle W}}

\newcommand{\hopc}{\kappa_{\rm c}}

\newcommand{\fP}{f_{\rm\scriptscriptstyle P}}

\newcommand{\fA}{f_{\rm\scriptscriptstyle A}}

\newcommand{\gX}{g_{\rm\scriptscriptstyle X}}
\newcommand{\gP}{g_{\rm\scriptscriptstyle P}}
\newcommand{\gS}{g_{\rm\scriptscriptstyle S}}
\newcommand{\gA}{g_{\rm\scriptscriptstyle A}}

\newcommand{\ZP}{Z_{\rm\scriptscriptstyle P}}
\newcommand{\ZS}{Z_{\rm\scriptscriptstyle S}}

\newcommand{\zf}{z_{\rm f}}
\newcommand{\ZPSF}{Z_{\rm\scriptscriptstyle P}^{\rm\scriptscriptstyle SF}}
\newcommand{\ZPchiSF}{Z_{\rm\scriptscriptstyle P}^{\rm\scriptscriptstyle \chi SF}}
\newcommand{\ZSchiSF}{Z_{\rm\scriptscriptstyle S}^{\rm\scriptscriptstyle \chi SF}}
\newcommand{\sigmaP}{\sigma_{\rm\scriptscriptstyle P}}
\newcommand{\sigmaS}{\sigma_{\rm\scriptscriptstyle S}}

\newcommand{\SigmaP}{\Sigma_{\rm\scriptscriptstyle P}}
\newcommand{\SigmaS}{\Sigma_{\rm\scriptscriptstyle S}}
\newcommand{\SigmaPS}{\Sigma_{\rm\scriptscriptstyle P/S}}
\newcommand{\SigmaSP}{\Sigma_{\rm\scriptscriptstyle S/P}}

\newcommand{\SigmaPSF}{\Sigma_{\rm\scriptscriptstyle P}^{\rm\scriptscriptstyle SF}}
\newcommand{\SigmaPchiSF}{\Sigma_{\rm\scriptscriptstyle P}^{\rm\scriptscriptstyle \chi SF}}
\newcommand{\SigmaSchiSF}{\Sigma_{\rm\scriptscriptstyle S}^{\rm\scriptscriptstyle \chi SF}}

\newcommand{\icsw}{c_{\rm sw}}
\newcommand{\ict}{c_{\rm t}}
\newcommand{\icttil}{\tilde c_{\rm t}}

\newcommand{\icA}{c_{\rm\scriptscriptstyle A}}
\newcommand{\icT}{c_{\rm\scriptscriptstyle T}}
\newcommand{\icV}{c_{\rm\scriptscriptstyle V}}

\newcommand{\uSF}{u_{\rm\scriptscriptstyle SF}}
\newcommand{\uGF}{u_{\rm\scriptscriptstyle GF}}

\newcommand{\abar}{\kern1pt\overline{\kern-1pt a\kern-0.5pt}\kern1pt}

\newcommand{\cO}{{\cal O}}
\newcommand{\cP}{{\cal P}}
\newcommand{\cQ}{{\cal Q}}

\graphicspath{{figs/}}
\begin{document}

\title{
\begin{flushright}
{\rm CERN-TH-2021-218}
\end{flushright}
Nonperturbative running of the quark mass for $N_f=3$ QCD
       from the chirally rotated Schr\"odinger Functional}
 
\author{\begin{center}
     \includegraphics[height=2.0\baselineskip]{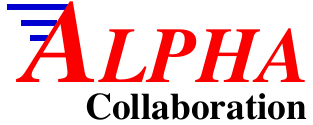}
       \end{center} Isabel~Campos~Plasencia}
\affiliation{Instituto de F\'isica de Cantabria IFCA-CSIC, Avda. de los Castros s/n, 39005 Santander, Spain}
\author{Mattia~Dalla~Brida}
\affiliation{Theoretical Physics Department, CERN, CH-1211 Geneva 23, Switzerland}
\author{Giulia~Maria~de~Divitiis}
\affiliation{Dipartimento di Fisica, Universit\`a di Roma ``Tor Vergata'', Via della Ricerca Scientifica 1, 00133 Rome, Italy}
\affiliation{INFN, Sezione di Tor Vergata, Via della Ricerca Scientifica 1, 00133 Rome, Italy}
\author{Andrew~Lytle}
\affiliation{Department of Physics, 
        University of Illinois at Urbana-Champaign,
        Urbana, Illinois, 61801, USA}
\author{Mauro~Papinutto}
\affiliation{Dipartimento di Fisica,  Universit\`a di Roma La Sapienza, and INFN, Sezione di Roma, Piazzale A.~Moro 2, 00185 Rome, Italy}
\author{Ludovica~Pirelli}
\affiliation{Dipartimento di Fisica, Universit\`a di Roma ``Tor Vergata'', Via della Ricerca Scientifica 1, 00133 Rome, Italy}
\affiliation{INFN, Sezione di Tor Vergata, Via della Ricerca Scientifica 1, 00133 Rome, Italy}
\author{Anastassios~Vladikas}
\affiliation{INFN, Sezione di Tor Vergata, Via della Ricerca Scientifica 1, 00133 Rome, Italy}
\date{\today}

\begin{abstract}
We study the Renormalisation Group (RG) running of the quark mass,
for $\NF=3$ QCD with Wilson fermions in a mixed action setup,
with standard Schr\"odinger Functional (SF) boundary conditions 
for sea quarks and  chirally rotated Schr\"odinger Functional 
($\chi$SF) boundary conditions for valence quarks. This necessitates
the tuning of the boundary factor $\zf(g_0^2)$ of the $\chi$SF
valence action, in order to ensure that QCD symmetries are fully
recovered in the continuum. The properties of this novel setup
are monitored through the ratios $\ZS/\ZP$ and $\SigmaS/\SigmaP$ of the renormalisation parameters and step 
scaling functions of the scalar and pseudoscalar densities. Where comparison is possible, our $\ZS/\ZP$ results
are found to agree with previous determinations, based on a mass ratio method~\cite{deDivitiis:2019xla} 
and Ward identities~\cite{Heitger:2020mkp,Heitger:2021bmg}, with  Schr\"odinger
Functional boundary conditions. The behaviour of $\SigmaS/\SigmaP$ confirms the theoretical  expectations of 
$\chi$SF QCD, related to the restoration of the theory's symmetries in the continuum limit. From the step scaling 
function of the pseudoscalar density we obtain the quark mass RG-running function from hadronic to perturbative energy scales.
This is fully compatible with the earlier result obtained in a similar setup for Wilson quarks with Schr\"odinger Functional boundary conditions~\cite{Campos:2018ahf}, and provides a strong universality test for the two lattice setups. 
\end{abstract}
\keywords{Lattice QCD; Renormalization group; Schr\"odinger functional; Chiral Symmetry restoration with Wilson fermions}
\pacs{12.38.Gc, 12.38.Aw, 11.10.Hi, 11.15.Ha}
\maketitle

\section{Introduction} 
\label{sec:intro}
The chirally rotated Schr\"odinger functional ($\chi$SF)~\cite{Sint:2010eh}
is a variant of the Schr\"odinger functional (SF) renormalisation scheme,
which enables us to 
obtain renormalisation parameters and lattice step scaling
functions, which are ``automatically'' 
$O(a)$-improved. 

At the formal level, continuum massless QCD with SF boundary
conditions (SF-QCD) is equivalent to the one with $\chi$SF boundaries ($\chi$SF-QCD),
as one is obtained from the other by suitable redefinitions
of the fermion fields, which are non-anomalous chiral rotations in isospin space.
These redefinitions also modify the form of the various symmetries
of the theory in the $\chi$SF setup. In particular, standard parity
$\cP$ in SF-QCD is chirally rotated to the so-called $\cP_5$ symmetry
in $\chi$SF-QCD.
When massless QCD is regularised  with Wilson fermions, the
strict equivalence between the theories with SF and $\chi$SF 
boundaries is lost. A consequence of this is that lattice QCD with SF
boundaries (SF-LQCD) retains $\cP$-symmetry, while its $\chi$SF 
counterpart ($\chi$SF-LQCD) is not $\cP_5$ symmetric.
The restoration of $\cP_5$ in the continuum limit requires the
introduction of a new counterterm at the time boundaries with a
coefficient $\zf(g_0^2)$. The new counterterm must be tuned by 
imposing that a $\cP_5$-odd correlation function in $\chi$SF-LQCD vanishes.
Once this is achieved (together with the tuning of the Wilson hopping
parameter to $\hopc$, which ensures that fermions are massless),
continuum limit universality implies that, at 
vanishing lattice spacing, $\cP_5$-even, renormalised  correlation
functions in $\chi$SF-LQCD are equal to their $\cP$-even counterparts
in SF-LQCD. The same is true for $\cP_5$-odd $\chi$SF-LQCD and $\cP$-odd 
SF-LQCD correlation functions; being pure lattice artefacts, these
vanish in the continuum.

In massless QCD with SF boundary conditions, $O(a)$-improvement is
achieved through the introduction of Symanzik counterterms in the action, 
both in the lattice bulk (i.e.\ the clover term of the fermion action with its coefficient $\icsw$) and at
the time boundaries (i.e.\ the terms with coefficients $\ict$ of the gauge action and $\icttil$ of
the fermionic one)~\cite{Luscher:1996sc}. Composite operators inserted in the lattice bulk in correlation 
functions also require their own counterterms (e.g.\ $\icA, \icV$ for the axial and vector
non-singlet currents respectively). The bulk improvement coefficients (e.g.\ $\icsw, \icA, \icV$)
are typically known nonperturbatively, whereas $\ict$ and $\icttil$ are known in 1- and sometimes in
2-loop perturbation theory. This implies that $O(a)$-effects are removed
in the bulk and $O(g_0^2 a)$-effects are removed from the time-boundaries. 
In practice, the former have always been found to be dominant, in the sense that
quantities thus improved scale like $O(a^2)$.

In massless QCD with $\chi$SF boundary conditions, bulk counterterms like,
$\icsw, \icA$ and $\icV$ are in principle not required for the removal of bulk $O(a)$
effects in the quantities of interest~\cite{Sint:2010eh} (``automatic improvement'').
The only necessary boundary counterterms are $\zf$ and $d_s$ (which are $O(a^0)$ 
and $O(a)$ respectively) for the fermionic action and $\ict$ (which is $O(a)$) for the 
gluonic one. We compute $\zf$ nonperturbatively, while $\ict$ is known at 1-loop in 
perturbation theory. For $d_s$ the tree-level value is adequate~\cite{Sint:2010eh,Brida:2016rmy};
this will be checked explicitly in this work.
This leaves us with  $O(a^2)$-effects
in the bulk and effectively with $O(g_0^4~a)$-effects at
the time-boundaries. Provided the latter turn out to be subdominant, the scaling of 
our results should be compatible with $O(a^2)$.

Several tests have been performed, which confirm this expectation:
\begin{itemize}
\item
In ref.~\cite{Sint:2010xy} ``automatic improvement'' was 
thoroughly investigated nonperturbatively in a quenched setup
for lattices with $L=1.436 r_0$,
with $r_0$ the Sommer scale~\cite{Sommer:1993ce}. The tuning
of $\zf$ was shown to be done correctly by checking the
vanishing of $\cP_5$-odd  $\chi$SF correlation functions in the 
continuum limit. Ratios of $\cP_5$-even $\chi$SF correlation functions
to their SF counterparts, which ought to have the same continuum limit,
are indeed shown to be compatible with unity, scaling like $O(a^2)$,
as they should.
\item
In ref.~\cite{Brida:2016rmy} analogous tests have been performed
in perturbation theory at 1-loop.
\item
A second quenched study~\cite{Lopez:2012as,Lopez:2012mc}
was centred on the step-scaling functions of the pseudoscalar
density and non-singlet twist-2 operators at one intermediate and
one perturbative scale. Moreover, the renormalisation factor
$\ZP$ was computed at the scale $L=1.436 r_0$.
Combining $\ZP$ with the bare twisted-mass parameter
and the known ratio $M/{\mbar(1/L)}$ of
ref.~\cite{Capitani:1998mq}, estimates of the strange quark mass
were obtained. Comparison of results
obtained with (i) unimproved SF Wilson fermions; (ii) SF clover fermions;
(iii) $\chi$SF fermions, confirmed
automatic improvement of the latter and the universality of the continuum
limit.
\item
In ref.~\cite{DallaBrida:2018tpn} these efforts have been extended to
massless QCD with $\NF=2$ and $\NF=3$ dynamical flavours. In the
$\NF=3$ case, the light sea quark doublet has been regularised in $\chi$SF-LQCD
and the third sea flavour in SF-LQCD.
\end{itemize}

A first goal of the present work is to pursue analogous tests from a different
angle. We use the gauge ensembles of ref.~\cite{Campos:2018ahf},
generated for the nonperturbative determination of the 
RG-running of the quark mass in $\NF=3$ massless QCD with SF boundaries.
The renormalisation scales $\mu$ cover a wide energy range 
$\lQCD \lessapprox \mu \equiv 1/L \lessapprox \MW$. Here $L$ denotes the
lattice physical extension; note that it ranges from very small values of about
$L \approx 10^{-3}$fm
to $L \lessapprox 1$fm.
On these ensembles we compute correlation functions with dimension-3 scalar 
and pseudoscalar   bilinear operators in the bulk and $\chi$SF boundary conditions. 
This is a mixed action approach, as sea and valence quarks have different regularisations
(SF and $\chi$SF respectively). So our whole setup is different to those
listed above. Several new universality tests are then possible, using different
$\chi$SF correlation functions which have the same continuum limit. 

The main result of this work consists of an estimate of $M/{\mbar(1/L)}$
and its comparison to the one obtained in a SF setup in ref.~\cite{Campos:2018ahf}.
This is a first step in a project, which ultimately aims at providing nonperturbative improved
estimates of the step scaling matrices of all four-fermion operators that contribute
to $B_{\rm\scriptscriptstyle{K}}$ in the Standard Model and beyond. This has already been
done for $\NF=2$, but in SF-LQCD~\cite{Dimopoulos:2007ht,Dimopoulos:2018zef}. 
Adopting a $\chi$SF-LQCD setup instead, brings in automatic $O(a)$-improvement of the
relevant dimension-6 operators and allows the introduction of renormalisation schemes
involving simpler correlation functions with better signal-to-noise behaviour.
The corresponding bare matrix elements are to be computed in 
twisted-mass LQCD at maximal twist (with Osterwalder-Seiler fermions), with the Clover term retained
in the action, for reasons explained in ref.~\cite{Dimopoulos:2009es}. For a first discussion
of this strategy, the reader may consult refs.~\cite{Mainar:2016uwb,Campos:2019nus}.

This paper is organised as follows: Sec.~\ref{sec:theory} is an overview of the fundamental properties
of the $\chi$SF lattice regularisation with Wilson quarks and the basic definitions and properties of the 
quantities of interest. Sec.~\ref{sec:simulation} describes the details of our numerical simulations.
The results of the tuning of the $\chi$SF-LQCD factor $\zf$, essential for the recovery of all QCD
symmetries in the continuum, is also presented.
In Sec.~\ref{sec:zratio} we obtain results for the ratio $\ZS/\ZP$ of the renormalisation factors
of the scalar and pseudoscalar bilinear operators. We find that the behaviour of $\ZS/\ZP$ at high energies
agrees with expectations from perturbation theory; at low energies it agrees with earlier nonperturbative
determinations, based on other methods. Moreover we confirm that the ratio $\SigmaSP$ of the corresponding
step scaling functions in the $\chi$SF theory goes to unity in the continuum limit, in accordance with
chiral symmetry restoration. In Secs.~\ref{sec:results-he} and \ref{sec:results-le} we obtain results for
the step scaling function of the pseudoscalar bilinear operator in the high- and low-energy regimes of our
simulations respectively. In each case we compute the RG running function
between the lowest and highest energy in each of the two regimes. Our results agree with those of 
ref.~\cite{Campos:2018ahf}. Finally, in Sec.~\ref{sec:concl} we conclude by presenting the total RG-running
factor between hadronic and very high (perturbative) scales. Several details of our analysis are treated
separately in the Appendices. Our results have been presented in preliminary form in
refs.~\cite{Campos:2019nus,Plasencia:2021gjp}.

\section{Theoretical Considerations} 
\label{sec:theory}
We now cover briefly those aspects of the $\chi$SF regularisation of refs.~\cite{Sint:2010eh,Brida:2016rmy} which are most relevant to our work. At the formal level, the massless QCD action is invariant under general flavour and chiral transformations.
In particular, it is invariant under the change of variables, 
\begin{equation}
\label{eq:ferm-rots}
\psi = R(\pi/2) \, \psi^\prime \,\, , \qquad  \bar \psi = \bar \psi^\prime \,  R(\pi/2) \,\, ,
\end{equation}
where $\psi, \bar \psi$ and $\psi^\prime, \bar \psi^\prime$ are doublets in isospin space
(e.g.\ $\psi = (\psi_u \,\, \psi_d)^T$), related through the above chiral non-singlet transformations 
with $R(\alpha) = \exp(i \alpha \gamma_5 \tau^3/2)$.
In the SF-QCD setup, lattices have finite physical volume $L^3 \times T$ (in the present work $T=L$),
with fields obeying Dirichlet (periodic) boundary conditions in time (space).
The former are defined at $x_0 = 0$ and $x_0 = T$ as follows:
\begin{eqnarray}
P_+ \psi(x) \big \vert_{x_0 = 0} = 0 \,\, , & \qquad & P_- \psi(x) \big \vert_{x_0 = T} = 0 \,\, ,
\nonumber \\
\bar \psi(x) P_-  \big \vert_{x_0 = 0} = 0 \,\, , & \qquad & \bar  \psi(x) P_+ \big \vert_{x_0 = T} = 0 \,\, ,
\end{eqnarray}
with projectors $P_\pm = (1 \pm \gamma_0)/2$. The chiral rotations (\ref{eq:ferm-rots}) map the above
conditions onto the $\chi$SF boundary conditions
\begin{eqnarray}
\tilde Q_+ \psi^\prime(x) \big \vert_{x_0 = 0} = 0 \,\, , & \qquad & \tilde Q_- \psi^\prime(x) \big \vert_{x_0 = T} = 0 \,\, ,
\nonumber \\
\bar \psi^\prime(x) \tilde Q_+  \big \vert_{x_0 = 0} = 0 \,\, , & \qquad & \bar  \psi^\prime(x) \tilde Q_- \big \vert_{x_0 = T} = 0 \,\, ,
\end{eqnarray}
with projectors $\tilde Q_\pm = (1 \pm i \gamma_0 \gamma_5 \tau^3)/2$. Thus SF-QCD and $\chi$SF-QCD 
are equivalent theories, since one is obtained from the other by the redefinition of fermionic fields~(\ref{eq:ferm-rots}).

Given the equivalence of the two theories, it is hardly surprising that they share all symmetries: the well known SF-QCD symmetries,
once transcribed in terms of fields $\psi^\prime$ and $\bar \psi^\prime$, are those of $\chi$SF-QCD. Flavour symmetry in
its standard SF-QCD version (e.g.\ Eq.~(2.15) of ref.~\cite{Brida:2016rmy}) takes the form of Eqs.~(2.16) and (2.17) 
of~\cite{Brida:2016rmy}; parity $\cP$ (Eq.~(2.18) of~\cite{Brida:2016rmy}) becomes $\cP_5$ (Eq.~(2.19) of~\cite{Brida:2016rmy}) 
in $\chi$SF-QCD. Charge conjugation is form-invariant in the two versions. We note in passing that the parity operator $\cP_5$ commutes with the boundary projectors $\tilde Q_\pm$.

Similar considerations apply to correlation functions. 
Following ref.~\cite{Brida:2016rmy}, we introduce, in $\chi$SF-QCD, a second flavour doublet 
$(\psi_{u^\prime} \,\, \psi_{d^\prime})^T$, with exactly the same properties as the original one. 
In SF-QCD the two point function with a pseudoscalar 
insertion $P^{ud}(x)$ in the bulk and a boundary operator $\cO_5^{du}$ at $x_0 = 0$ is defined as
\begin{equation}
\label{eq:fP_def}
\fP \,\,  = \,\, -\dfrac{1}{2} \langle P^{ud} \cO_5^{du} \rangle_{(P_+)} \,\, .
\end{equation}
Analogous quantities in $\chi$SF-QCD are
\begin{equation}
\label{eq:gX_def}
\gX^{f_1 f_2} \,\,  = \,\, -\dfrac{1}{2} \langle X^{f_1 f_2} \cQ_5^{f_2 f_1} \rangle_{(\tilde Q_+)} \,\, ,
\end{equation}
with $X^{f_1 f_2} = P^{f_1 f_2}$ or $X^{f_1 f_2} = S^{f_1 f_2}$, and $\cQ_5^{f_2 f_1}$ the result of the boundary field 
rotations~(\ref{eq:ferm-rots}) on the operator $\cO_5$. The allowed combinations of flavour indices are
$(f_1, f_2) = (u, u^\prime), (d, d^\prime), (u, d), (d, u)$ (so that no disconnected diagrams arise). 
See ref.~\cite{Brida:2016rmy} for more detailed explanations.
The relations between these correlation functions are then
\begin{equation}
\fP \, = \, i \gS^{uu^\prime} \, = \, -i \gS^{dd^\prime} \, = \,  \gP^{ud} \, = \, \gP^{du} \, \, .
\end{equation}
The SF boundary-to-boundary correlation function 
\begin{equation}
f_1 \, = \, -\dfrac{1}{2}  \langle \cO_5^{ud} \cO_5^{\prime du} \rangle_{(P_+)} 
\end{equation}
(with the boundary operator $\cO_5^{\prime du}$ defined at $x_0 = T$) and  its $\chi$SF counterpart
\begin{equation}
g_1^{f_1f_2} \, = \, -\dfrac{1}{2}  \langle \cQ_5^{f_1f_2} \cQ_5^{\prime f_2 f_1} \rangle_{(\tilde Q_+)} 
\end{equation}
are also related:
\begin{equation}
f_1 \, = \, g_1^{uu^\prime} \, = \, g_1^{dd^\prime} \, =  \, g_1^{ud} \, =  \, g_1^{du} \, \ .
\end{equation}

The above properties, though trivial at the formal level, have non-trivial consequences once the lattice
regularisation with Wilson fermions ($\chi$SF-LQCD) is introduced. (Of the three lattice $\chi$SF-QCD versions
proposed in ref.~\cite{Sint:2010eh}, we use that of ref.~\cite{Brida:2016rmy}; see Sec.~3.1 of the latter
work for the definition of the action etc.). The Wilson term and boundary terms in $\chi$SF-LQCD
induce the breaking of the rotated flavour symmetry (i.e.\ Eqs.~(2.16) and (2.17) of~\cite{Brida:2016rmy})
and parity $\cP_5$ (i.e.\ Eq.~(2.19) of~\cite{Brida:2016rmy}). However, a symmetry argument analogous to that introduced
in twisted-mass QCD~\cite{Frezzotti:2003ni} holds in the present case~\cite{Sint:2007ug,Sint:2010eh}, with the
result that $\cP_5$-even correlation functions of the $\chi$SF-LQCD theory, once renormalised, are
$O(a)$-improved in the bulk. An important additional ingredient of the lattice formulation consists in the introduction of
the following boundary terms in the action: 
\begin{eqnarray}
\label{eq:xhiSFcounterterms}
&& \bar \psi(x) [ \delta D_{\rm\scriptscriptstyle W} ] \psi(x) = (\delta_{x_0,0} + \delta_{x_0,T}) \nonumber \\
&& \times \bar \psi(x) \big [(\zf-1) + (d_s-1) a {\bf D}_s \big ] \psi(x) \,\, .
\end{eqnarray}
The operator ${\bf D}_s$ is the Dirac-Wilson lattice operator, summed over the three spatial directions
only (see Eq.~(3.14) of ref.~\cite{Brida:2016rmy}). It is an improvement counterterm, which cancels boundary
$O(a)$-effects, once the coefficient $d_s(g_0^2)$ is properly tuned. 
Moreover, the aforementioned symmetry-breaking pattern of $\chi$SF-LQCD 
necessitates the introduction of an additional $O(a^0)$ boundary operator with coefficient $\zf(g_0^2)$,
which must be appropriately tuned, 
in order for the rotated flavour and $\cP_5$ symmetries to be recovered
in the continuum. The tuning condition consists in finding, at finite lattice spacing (i.e.\ non-zero $g_0^2$)
the value of $\zf$ for which a $\cP_5$-odd correlation function vanishes. Specifically we require that
\begin{equation}
\label{eq:tune-zf}
\gA^{ud}(x_0) \Big \vert_{x_0 = T/2} \,\, = \,\, 0 \,\, ,
\end{equation}
where the $\cP_5$-odd $\gA^{ud}$ is defined in Eq.~(\ref{eq:gX_def}), with $X=A_0$. The above tuning must be coupled to the requirement that the theory be massless; i.e.\ the hopping 
parameter $\kappa$ must be tuned to its critical value $\hopc$.  This can be achieved 
by requiring the vanishing of the current quark mass
\begin{equation}
\label{eq:tune-kappac}
m(g_0^2,\hopc)  \Big \vert_{x_0 = T/2} \,\, = \,\, 0 \,\, ,
\end{equation}
which in $\chi$SF-LQCD may be defined as~\cite{DallaBrida:2018tpn}
\begin{equation}
\label{eq:m-PCAC-chiSF}
m^{\rm\scriptscriptstyle \chi SF}(g_0^2,\kappa) \, \equiv \, \dfrac{\dfrac{1}{2}(\partial_0 + \partial^\ast_0) \gA^{ud}(x_0)}{2   \gP^{ud}(x_0)} \,\, ,
\end{equation}
where  $\partial_0, \partial^\ast_0$ are forward and backward lattice derivatives respectively. Recall that in SF-LQCD, in standard ALPHA fashion, the current quark mass is defined by~\cite{Capitani:1998mq}
\begin{equation}
\label{eq:m-PCAC}
m^{\rm\scriptscriptstyle SF}(g_0^2,\kappa) \, \equiv \, \dfrac{\dfrac{1}{2}(\partial_0 + \partial^\ast_0) \fA(x_0) + a \icA \partial^\ast_0 \partial_0 \fP(x_0)}
{2  \fP(x_0)} \,\, .
\end{equation}
In the above $\fA$ is the analogue of Eq.~(\ref{eq:fP_def}), with $A_0^{ud}$ replacing $P^{ud}$. In SF-QCD, $\kappa$ is tuned to its critical value $\hopc$ by requiring the vanishing of $m^{\rm\scriptscriptstyle SF}$.

The SF and $\chi$SF renormalisation conditions for the pseudoscalar and scalar operators are imposed in the usual manner~\cite{Capitani:1998mq,Brida:2016rmy}
\begin{eqnarray}
\dfrac{\ZPSF(g_0^2,L/a) \fP(T/2)}{\sqrt{f_1}} &=& \Bigg [ \dfrac{\fP(T/2)}{\sqrt{f_1}} \Bigg ]^{\rm \scriptstyle t.l.} \,\,, \label{eq:ZP-SF} \\
\dfrac{\ZPchiSF(g_0^2,L/a) \gP^{ud}(T/2)}{\sqrt{g_1^{ud}}} &=& \Bigg [ \dfrac{\gP^{ud}(T/2)}{\sqrt{g_1^{ud}}} \Bigg ]^{\rm \scriptstyle t.l.} \,\,,\label{eq:ZP-chiSFP} \\
\dfrac{\ZSchiSF(g_0^2,L/a) \gS^{uu^\prime}(T/2)}{\sqrt{g_1^{uu^\prime}}} &=& \Bigg [ \dfrac{\gS^{uu^\prime}(T/2)}{\sqrt{g_1^{uu^\prime}}} \Bigg ]^{\rm \scriptstyle t.l.} \,\,,\label{eq:ZP-chiSFS} 
\end{eqnarray}
where the superscripts $\rm \scriptstyle t.l.$ on the r.h.s. stand for ``tree level'' (these tree level quantities are computed at non-vanishing $a/L$). From them we determine the renormalisation constants $\ZP$ in the SF and $\chi$SF renormalisation schemes and $\ZS$ in the $\chi$SF  scheme. 

As a side remark we point out that standard parity $\cP$, combined with flavour exchanges $u \leftrightarrow d$ and  $u^\prime \leftrightarrow d^\prime$ is an exact 
symmetry of $\chi$SF-LQCD~\cite{Sint:2010eh,Brida:2016rmy}. This ensures that $\gS^{uu^\prime} = - \gS^{dd^\prime}$ and $\gP^{ud} = \gP^{du}$ are exact lattice relations. For this reason we have not used the correlation functions $ \gS^{dd^\prime}$ and $\gP^{du}$, as they do not convey any new information.

The definitions of the lattice step scaling functions (SSF) are also standard:
\begin{eqnarray}
\SigmaPSF(g_0^2,a/L) &=& \dfrac{\ZPSF(g_0^2,2L/a)}{\ZPSF(g_0^2,L/a)} \,\, , \label{eq:Ssf-PSF} \\
\SigmaPchiSF(g_0^2,a/L) &=& \dfrac{\ZPchiSF(g_0^2,2L/a)}{\ZPchiSF(g_0^2,L/a)} \,\, , \label{eq:Ssf-PchiSF} \\
\SigmaSchiSF(g_0^2,a/L) &=& \dfrac{\ZSchiSF(g_0^2,2L/a)}{\ZSchiSF(g_0^2,L/a)} \,\, . \label{eq:Ssf-SchiSF}
\end{eqnarray}
On the lattice, SF-QCD and $\chi$SF-QCD are two regularisations of the same continuum theory, in which the pseudoscalar and scalar operators belong to the same symmetry multiplets (such as the chiral multiplet) and thus have the same anomalous dimension. Consequently, the above SSFs should have the same continuum limit:
\begin{equation}
\sigmaP(u) \,\, = \,\, \lim_{a \rightarrow 0} \Sigma_{\rm\scriptscriptstyle X}^{\rm\scriptscriptstyle Y}(g_0^2,a/L) \Big \vert_{\gbar^2(L) = u} \,\, .
\end{equation}
In the above (X,Y) = (P, SF), (P, $\chi$SF), (S, $\chi$SF). The squared renormalised coupling $\gbar^2(L)$ is meant to be held fixed at a value $u$ while the continuum limit is taken. In terms of the renormalised quark mass $\mbar(\mu)$, defined at a scale $\mu=1/L$, which corresponds to a renormalised coupling $\gbar^2(\mu) = u$, the continuum step scaling function is given by the ratio
\begin{equation} \label{sigmaP-mbar}
\sigmaP(u) \,\, = \,\,\dfrac{\mbar(\mu)}{\mbar(\mu/2)} \,\,  .
\end{equation}

So far we have dealt with SF-QCD and $\chi$SF-QCD as two distinct, if related, setups. As noted in ref.~\cite{Sint:2010eh}, for an odd number of flavours, the fermion determinant in the $\chi$SF formalism is in general complex. For $\NF=3$ QCD the problem has been circumvented in ref.~\cite{DallaBrida:2018tpn} by working with a $\chi$SF-LQCD light sea quark doublet and a SF-LQCD third sea flavour. Here we adopt a different mixed action approach, with the sea quark action obeying standard SF boundary conditions, and the valence quark action defined for an even number of fermions obeying  $\chi$SF boundary conditions.  For the sea quarks, we use the existing SF-QCD configuration ensembles of ref.~\cite{Campos:2018ahf}. The novelty with respect to ref.~\cite{Campos:2018ahf} thus consists in having valence fermions organised in doublets with $\chi$SF boundary conditions. Apart from this, the lattice gauge action, the fermion action in the lattice bulk (Wilson fermions with a clover term), and the renormalised coupling definition(s) remain the same. Since our lattice valence quark propagators are now computed in a $\chi$SF setup, it is $\chi$SF symmetries that determine the renormalisation and improvement properties of the correlation functions and the quantities derived from them. Thus we expect $O(a)$-improvement to be ``automatic" in the bulk (i.e.\ in general bulk operators do not require Symanzik coefficients, e.g.\ $\icA$ and $\icT$ for the axial and tensor bilinears).

In this setup we use renormalisation conditions~(\ref{eq:ZP-chiSFP},\ref{eq:ZP-chiSFS}) for the computation of $\ZP$ and $\ZS$ and definitions~(\ref{eq:Ssf-PchiSF},\ref{eq:Ssf-SchiSF}) for the SSFs. The three SSFs, computed from \eqref{eq:Ssf-PSF} in ref.~\cite{Campos:2018ahf} and Eqs.~(\ref{eq:Ssf-PchiSF}), (\ref{eq:Ssf-SchiSF}) in this work, should yield the same SSF $\sigmaP$ in the continuum, since the same renormalisation conditions are imposed. 

In a purely $\chi$SF-LQCD setup, the necessary tuning of $\zf$ is based on Eq.~(\ref{eq:tune-zf}), while that of $\kappa$ on Eqs.~(\ref{eq:tune-kappac}) and (\ref{eq:m-PCAC-chiSF}). In practice the tuning of the two parameters can be done independently, as they depend weakly on each other~\cite{DallaBrida:2018tpn}. In our mixed action setup, we avoid tuning $\kappa$ altogether, as we can use the $\hopc$ results of ref.~\cite{Campos:2018ahf}, which are based on Eq.~(\ref{eq:m-PCAC}). Moreover, the tuning of 
$\zf$ is performed exclusively in the valence sector, given that our sea SF-QCD quarks are ``blind" to this factor. This is to be contrasted to the much costlier $\zf$ tuning in the purely $\chi$SF-LQCD case, where the generation of the gauge ensembles depends on $\zf$.

\section{Numerical Simulations} 
\label{sec:simulation}
We have seen in Sec.~\ref{sec:theory} that sea quarks are regularised as 
explained in ref.~\cite{Campos:2018ahf}. For this reason, the first
part of this section consists in a recapitulation of aspects of
that work which are relevant to the present one. 

The lattice volumes $L^4$ in which simulations are performed
define the range of accessible energy scales $\mu = 1/L$. 
Essentially there are two energy regimes. The 
high-energy one concerns scales in the range $\mu_0/2 \lessapprox \mu
\lessapprox \MW$, with an intermediate (``switching") scale conventionally
chosen to be $\mu_0/2 \sim 2$~GeV. The low-energy regime
concerns scales in the range $\lQCD \lessapprox \mu
\lessapprox  \mu_0/2$. The main difference between the 
two~\cite{Brida:2016flw,DallaBrida:2016kgh,DallaBrida:2018rfy} is the definition adopted 
for the renormalised coupling $\gbar$: in the high-energy range it is
the nonperturbative SF coupling first introduced in ref.~\cite{Luscher:1992an,Luscher:1993gh};
in the low energy one it is the gradient flow (GF) coupling defined in
ref.~\cite{Fritzsch:2013je}. This allows to optimally exploit the variance properties
of the couplings, so that a very precise computation of the 
Callan-Symanzik $\beta$-function and ultimately of $\lQCD$ 
is achieved~\cite{Bruno:2017gxd}.

In refs.~\cite{Brida:2016flw,DallaBrida:2016kgh,DallaBrida:2018rfy},
the switching scale, $\mu_0/2$, where we switch between the SF and GF definitions
of the coupling, was given implicitly by the formula:
\begin{equation}
\gbar_{\SF}^2(\mu_0) \equiv \uSF(\mu_0) = 2.0120 \,,
\end{equation}
corresponding to the largest value for the renormalised coupling
on the SF ensembles used in the analysis.
The value of the SF coupling was determined down to the scale $\mu_0/2$
in~\cite{Brida:2016flw}; this amounts to computing the SSF of the SF coupling
\begin{equation}
\sigma_{\SF}(u_0) \equiv \gbar_{\SF}^2(\mu_0/2) = \uSF(\mu_0/2) = 2.452(11) \,\,.
\end{equation}
The matching between schemes was subsequently specified by determining
the value of the GF coupling at the same scale~\cite{DallaBrida:2016kgh}:
\begin{equation}
\gbar_{\GF}^2(\mu_0/2) = \uGF(\mu_0/2) = 2.6723(64) \,\, .
\end{equation}
In physical units this corresponds to a switching scale $\mu_0/2$ of 
approximately 2 GeV~\cite{Bruno:2017gxd}.

Moreover, different lattice regularisations were adopted in each energy
regime. At high energies, simulations were carried out 
using the plaquette gauge action~\cite{Wilson:1974sk}
and the clover fermion action~\cite{Sheikholeslami:1985ij} 
with the nonperturbative value of  $\icsw$~\cite{Yamada:2004ja} and
the one-~\cite{Sint:1997jx} and two-loop~\cite{Bode:1998hd}
values of $\icttil$ and $\ict$ respectively. At low energies the tree-level Symanzik 
improved (L\"uscher-Weisz) gauge action was
used~\cite{Luscher:1985zq}. The fermion action was the $O(a)$-improved 
clover~\cite{Sheikholeslami:1985ij}, with the nonperturbative value 
of the improvement coefficient $\icsw$~\cite{Bulava:2013cta} and 
one-loop values of $\icttil$~\cite{VilasecaPrivate,DallaBrida:2016kgh} and $\ict$~\cite{Aoki:1998qd}.

Note that in ref.~\cite{Campos:2018ahf} the same SF 
renormalisation condition was used in both energy regimes for 
the determination of the quark mass renormalisation factor 
$1/\ZP$, its step scaling function etc. This implies that $\sigmaP$ and $\mbar$  are expected to be 
continuous functions of the renormalisation scale $\mu$ in the whole simulation range $[\lQCD, \MW]$. 
The same quantities, when plotted against the squared renormalised coupling $u$, 
will be discontinuous at the $u$-value corresponding to the switching scale $\mu_0/2$, 
due to different definitions of the coupling below and above this scale. 
Any quantity is also going to be a discontinuous function of 
the squared inverse bare coupling $\beta$, as the bare actions are different in the two regimes.

At high energies, simulations were carried out \cite{Campos:2018ahf} for 8 values of the
squared renormalised coupling $\uSF$. For each of these couplings, 
corresponding to a fixed renormalisation scale $\mu = 1/L$, the inverse 
bare coupling $\beta = 6/g_0^2$ was tuned appropriately for
$L/a = 6,8,12$ ($a$ is the lattice spacing).
At the strongest coupling $\uSF = 2.012$ of this high-energy range,
an extra finer lattice with $L/a = 16$
was simulated. At low energies, simulations were carried out for 7
values of the squared renormalised coupling $\uGF$. The inverse bare coupling $\beta = 6/g_0^2$ 
was chosen so that $\uGF$ remains {\it approximately} constant for the three lattice volumes $L/a = 8,12,16$.
In both the high and the low energy ranges, gauge ensembles were generated at 
each $(\beta,L/a)$ and $(\beta,2L/a)$ combination. At fixed $(\beta,L/a)$ the 
hopping parameter $\kappa$ was tuned to its critical value $\hopc$, so that the bare 
$O(a)$-improved PCAC mass defined in Eq.~(\ref{eq:m-PCAC}) vanishes at the
corresponding value of $\beta$; cf.\ Eq.~(\ref{eq:tune-kappac}).
For each $(\beta,L/a,\hopc)$ and $(\beta,2L/a,\hopc)$ the factors $\ZPSF(g_0^2,L/a)$
and $\ZPSF(g_0^2,2L/a)$ were computed using Eq.~(\ref{eq:ZP-SF}). Their ratio gives
the SSF $\SigmaPSF$  defined in Eq.~(\ref{eq:Ssf-PSF}). More details can be found in 
ref.~\cite{Campos:2018ahf} and the Tables~6,~9  (SF range) and~8,~10 (GF range)
of that work.

\begin{figure}
\includegraphics[width=0.5\textwidth]{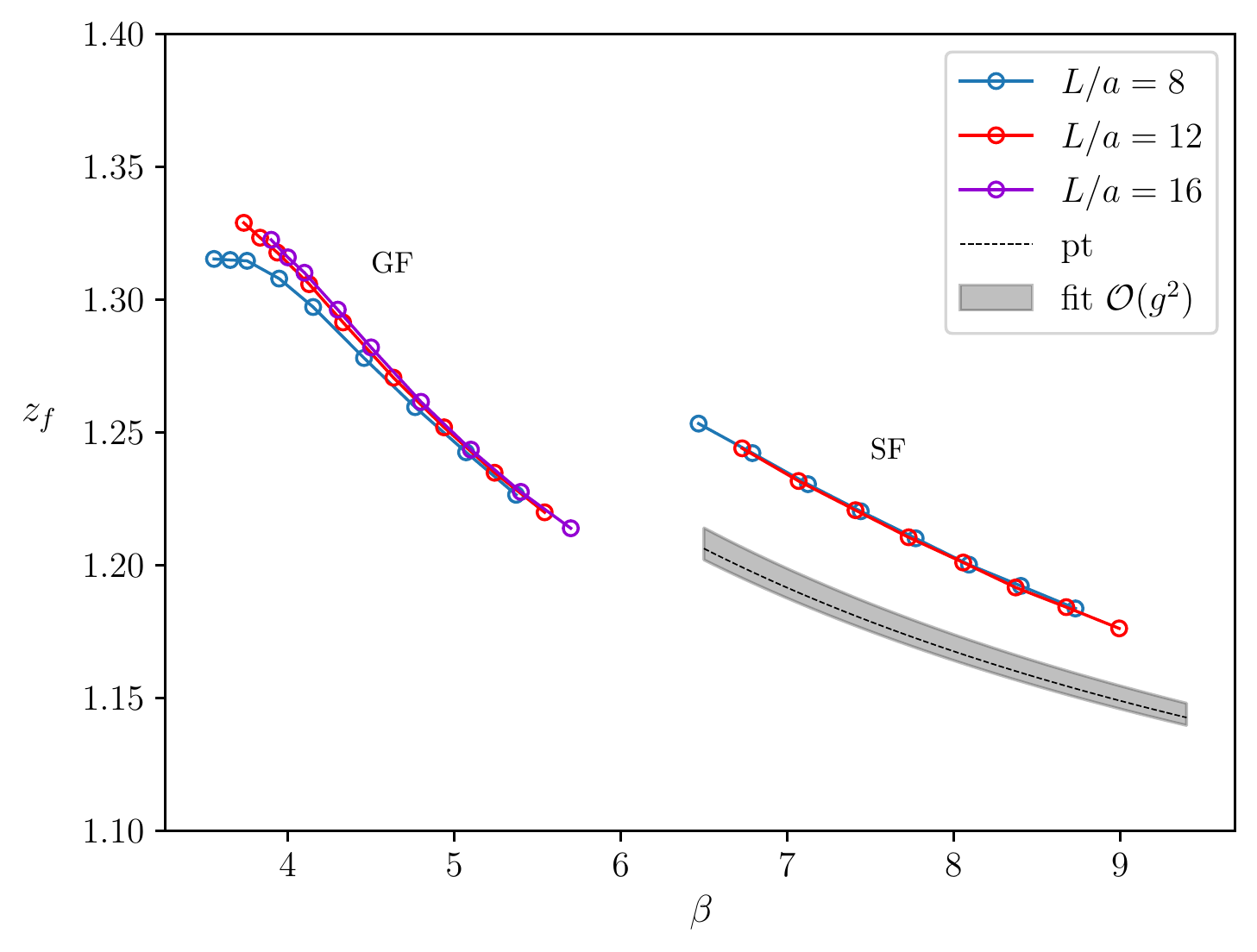}
\caption{The boundary counterterm $\zf$ on the high
energy (SF) and low energy (GF) ensembles. The black dotted line is the
perturbative result known at $O(g_0^2)$. The grey band is the
result from fitting the SF data, plotted after the resulting expression
has been truncated at $O(g_0^2)$ (i.e.\ the
fit to the data correctly reproduces the perturbative prediction).
\label{fig:zf_vs_beta}}
\end{figure}

So far we have recapitulated the simulations of ref.~\cite{Campos:2018ahf}, performed in
$\NF=3$ QCD with sea and valence quarks subjected to SF boundary conditions. In the 
present work, we use the same configuration ensembles, with the exception of some $\beta$'s
where a few subsets of configurations could not be recovered; no significant loss in
statistical accuracy resulted from this. 
We invert the Dirac-Wilson operator with $\chi$SF boundary conditions.
Consequently $\zf$ has to
be determined so that Eq.~\eqref{eq:tune-zf} is satisfied. The results of $\zf$ as a function of $\beta$
for both high- and low-energy regimes are displayed in Fig.~\ref{fig:zf_vs_beta}. It should be stressed
that determining $\zf$ is essential in order to ensure that in our mixed action approach
chiral and flavour symmetries are recovered in the continuum and thus the theory belongs to the same
universality class as other lattice regularisations. This is corroborated by the results of Sec.~\ref{sec:zratio}.

Details of the tuning procedure leading to $\zf$ are discussed in Appendix~\ref{sec:zf}. As stated in Sec.~\ref{sec:theory}, the tuning of $\zf$ must in principle be coupled with that of $\kappa$ for the theory to be massless. This is not so in the procedure outlined above and in Appendix~\ref{sec:zf}, where we used the $\hopc$ estimates of ref.~\cite{Campos:2018ahf}, based on the SF quark mass definition of Eq.~\eqref{eq:m-PCAC}. In the $\chi$SF setup the PCAC quark mass, defined by Eq.~\eqref{eq:m-PCAC-chiSF}, is not expected to be exactly zero when $\hopc$ is tuned in SF-LQCD. 
(The difference however is an $O(a)$ cutoff effect which induces $O(a^2)$ corrections in $\cP_5$-even quantities.) 
One would expect that an iterative procedure in which $\zf$ and $\kappa$ are alternatively tuned would be needed. Such a procedure is adopted in Appendix~\ref{sec:kappa}, where it is demonstrated that the tuning of $\hopc$ alongside that of $\zf$ is not necessary in practice.

The counterterm $d_s$, introduced in Eq.\eqref{eq:xhiSFcounterterms}, is known at tree level~\cite{Sint:2010eh}
and, for the plaquette action, also at 1-loop order~\cite{Brida:2016rmy}:
\begin{equation}
d_{s}=\frac{1}{2}+d_{s}^{(1)}g_0^2 \,\, ,
\end{equation}
where
\begin{equation}
\label{eq:ds1}
d_{s}^{(1)}=-0.0006(3)\times C_{F} \,\, .
\end{equation}
For the L\"uscher-Weisz action $d_{s}^{(1)}$ is not known at present.
In  Appendix~\ref{sec:ds_check} it is explicitly shown that results are unaffected when the 1-loop estimate of $d_{s}$
is used instead of its tree level value.

For global fits throughout this work we use the
lsqfit~\cite{lsqfit} and gvar~\cite{gvar}
packages for correlated fitting and error propagation.
We have checked
that these results are consistent with jackknife and the $\Gamma$-method error analysis of ref.~\cite{Wolff:2003sm}. 
Except for the last method, data have been binned by 20 configurations. The code to compute the $\chi$SF correlation functions is built on openQCD 1.0 and previously  used in ref.~\cite{DallaBrida:2018tpn}. 

In order to facilitate future use of our $\chi$SF/SF-LQCD setup, we have gathered all relevant simulation details in Appendix~\ref{app:zfTables}. These are the number of measurements $N_{\rm\scriptscriptstyle ms}$ for each ensemble, the lattice size $L/a$, the bare parameters $\beta, \hopc$, the action coefficients $\icsw, \zf$, the renormalised squared coupling $u$, and the functions $\gA^{ud}$ and $\partial \gA^{ud}/\partial \zf$ used for the tuning of $\zf$.
\section{The ratio $\ZS/\ZP$}
\label{sec:zratio}

Now we turn to the ratio of renormalisation factors $\ZS/\ZP$.
In Wilson formulations of lattice gauge theory, 
it is a finite quantity that depends on the bare gauge coupling,
\begin{equation}
\label{eq:ZPZS}
\frac{\ZS}{\ZP} \simeq 1 + \sum_{i=1,j=0}^\infty c_{ij} g_0^{2i} 
\Big (\frac{a}{L} \Big)^{2j} \,.
\end{equation} 
In the $g_0 \rightarrow 0$ limit the above expression would be written 
as an equality, if terms containing products like
$[a/L]^{2j} [\ln(a/L)]^{k}$ ($j > 0, k > 1$)
were also added to the series; cf.\ Eq.(7.4) of ref.~\cite{Brida:2016rmy}.
We drop these terms, which are habitually neglected, as they cannot be resolved by the
data. We have also ignored nonperturbative contributions depending on $a \lQCD$.
We use the above expression for analysing $\ZS/\ZP$ in the high-energy region.
At a fixed bare coupling there is a finite $a/L \to 0$ limit,
and for our lattice setup the leading coefficient $c_{10}$ has been calculated
in lattice perturbation theory, $c_{10} = 0.025944(3)$~\cite{Brida:2016rmy}.

Unlike the renormalisation factors $\ZS$ and $\ZP$ themselves, their ratio does
not depend on the scale $\mu$. Its continuum limit is known to be unity. Moreover,
it has been calculated by other methods in the low energy range for our lattice setup.
Therefore it is suitable for some crosschecks of our results.
Following ref.~\cite{Brida:2016rmy}, we compute $\ZS/\ZP$ from the ratio
\begin{equation}
R^g_{\rm\scriptscriptstyle SP} \equiv 
\dfrac{g^{ud}_{\rm\scriptscriptstyle P}(T/2)}
{i g^{uu^\prime}_{\rm\scriptscriptstyle S}(T/2)} = 
\dfrac{\ZS}{\ZP} + O\Big(\dfrac{a^2}{L^2}\Big) \,\, .
\end{equation}
Note that $O(a)$ boundary effects cancel in this ratio, leaving us with $O(a^2)$ uncertainties.

\begin{figure}[ht]
\includegraphics[width=0.5\textwidth]{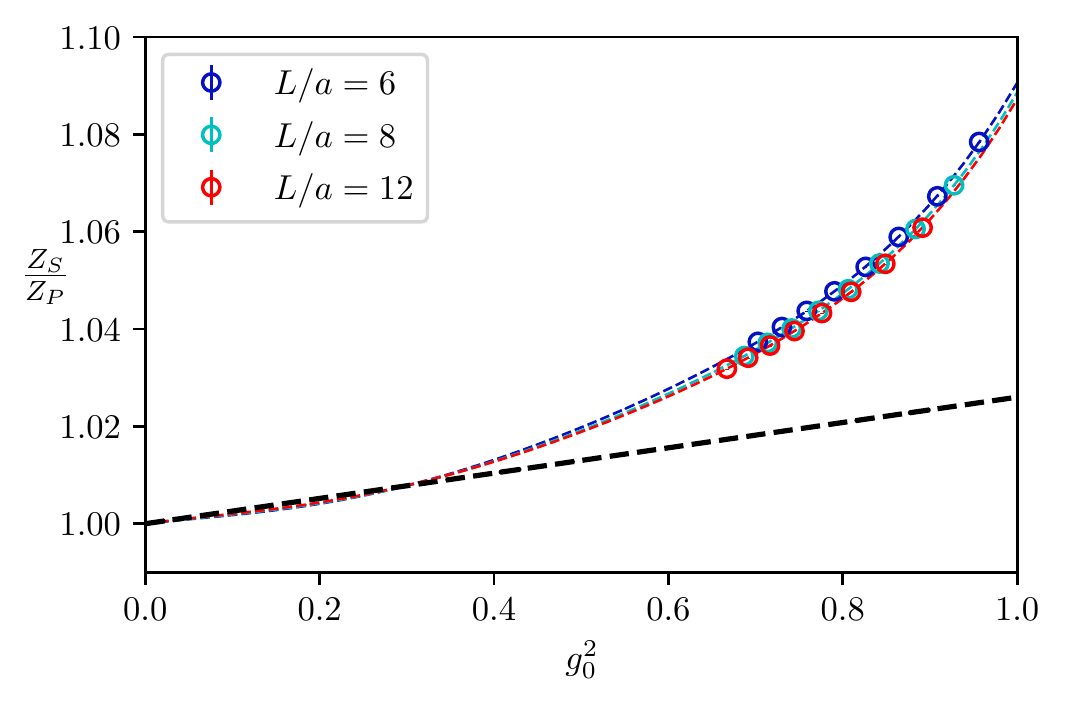}
\caption{The ratio $\ZS / \ZP$ in the high energy (SF) regime. 
The coloured circles show our numerical data, and the coloured
dotted lines
show the results of the fit to Eq.~\eqref{eq:ZPZS}
evaluated at the respective
$L/a$ values.
The black dotted
line gives the $O(g_0^2)$ perturbative result.
\label{fig:ZS_over_ZP}}
\end{figure}

In Fig.~\ref{fig:ZS_over_ZP} we show data for $\ZS/\ZP$ in the high energy regime,
where contact is made with perturbation theory. Due to the high degree of 
statistical correlation between $g^{ud}_{\rm\scriptscriptstyle P}(x_0)$ and
$g^{uu^\prime}_{\rm\scriptscriptstyle S}(x_0)$,  the error bars are extremely small, 
of order $10^{-5}$ (as compared to $10^{-3}$ for the correlators individually).
Nevertheless we are able to fit the data, 
performing global fits according to 
Eq.~\eqref{eq:ZPZS}, provided enough terms in $g_0^2$ are retained.
The coefficient $c_{10}$ is kept fixed at its 
perturbative value and the series is truncated at $(i,j) = (5,2)$. The goodness
of the fit is $\chi^2/{\rm d.o.f.} \sim 0.68$. Thus the numerical results appear
to be joining smoothly onto the one-loop perturbative result at  small bare
coupling. If the term $c_{10}$ is allowed to vary, the fit returns $c_{10} = 0.073(57)$
and $\chi^2/{\rm d.o.f.} = 0.66$, compatible at $1 \sigma $ with
the perturbative value, but only weakly constrained.
The three fit results for fixed $L/a=6,8,12$ are shown as dashed lines in Fig.~\ref{fig:ZS_over_ZP}.
Although they lie extremely close to each other, they are clearly distinct
curves. These differences imply that finite size effects are tiny.
This analysis has not been carried out along lines of constant physics;
$O(a^2 \lQCD^2)$ effects have been neglected. It therefore probes
the validity of perturbative expectations in a wide range of high energy scales.

\begin{figure}[ht]
\includegraphics[width=0.5\textwidth]{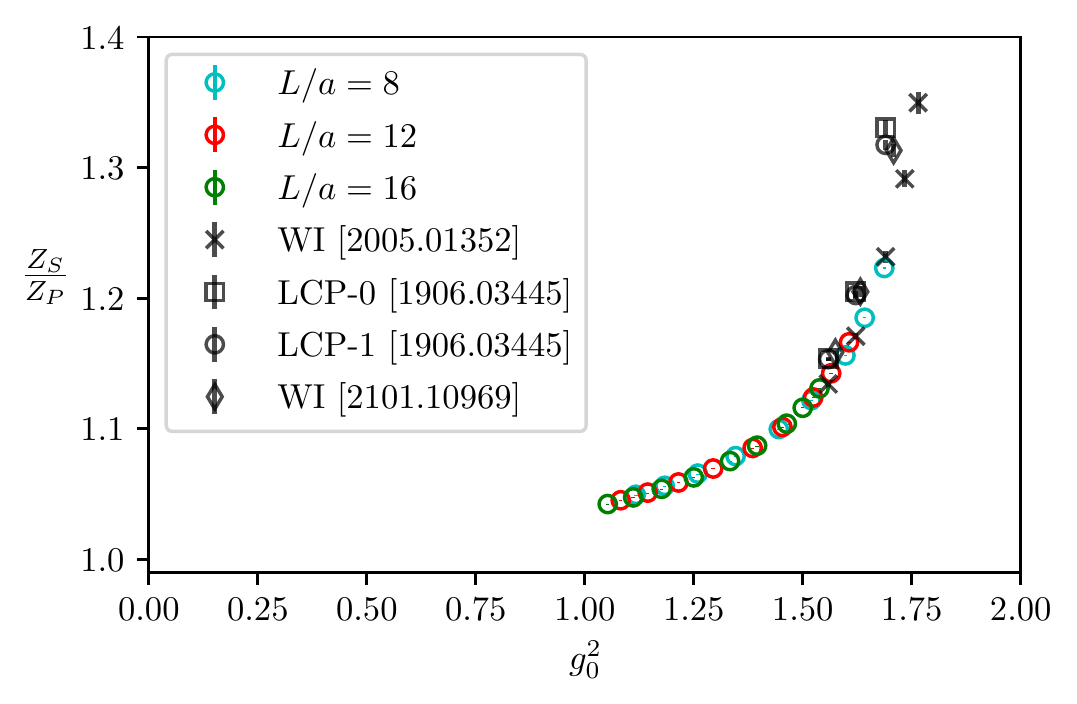}
\caption{The ratio $\ZS / \ZP$ in the low energy (GF) regime. Data is compared
with the results of~\cite{deDivitiis:2019xla} using quark-mass ratio methods
(LCP-0,1) 
and~\cite{Heitger:2020mkp,Heitger:2021bmg} obtained from Ward identities (WI).
\label{fig:ZS_over_ZP-gf}}
\end{figure}

At low-energies (GF) we show results for $\ZS/\ZP$ in 
Fig.~\ref{fig:ZS_over_ZP-gf}, and compare them
with recent results obtained in~\cite{deDivitiis:2019xla} 
from suitable ratios of current and
subtracted quark masses at two lines of constant physics (LCP-0,1) 
and in~\cite{Heitger:2020mkp,Heitger:2021bmg} using 
Ward Identities (WI). These works
use the same bulk action as the present one, with Schr\"odinger functional 
boundary conditions; quark masses lie close to the chiral limit; 
gauge couplings straddle the range of values of CLS simulations~\cite{Bruno:2014jqa,Mohler:2017wnb}
suitable for the computation of low-energy hadronic quantities.
Our results are compatible with those of the other methods
in the infrared. The comparison of results from~\cite{deDivitiis:2019xla} 
and~\cite{Heitger:2020mkp,Heitger:2021bmg} was discussed
already in~\cite{Heitger:2020mkp,Heitger:2021bmg}, and it was observed that the WI
method has smaller lattice artefacts. This can be expected to
translate to an improved control of the continuum extrapolation of
the quantities requiring $\ZS/\ZP$. 
Our results feature almost no visible finite volume effects 
and coincide numerically with the ones from~\cite{Heitger:2020mkp} 
in the region of $g_0^2$ where they overlap. At $g_0^2$ values larger than 
about 1.6, different methods give slightly different results, 
signalling the presence of sizeable discretisation effects.

Besides the ratio $\ZS / \ZP$, we also examine the ratio of the corresponding
SSF's $\SigmaPS \equiv \SigmaP/\SigmaS$. This is also a scale-independent
quantity which becomes unity in the continuum limit. Once again, $O(a)$ boundary 
effects cancel in this ratio, leaving us with $O(a^2)$ uncertainties.
This quantity is particularly suitable for studying universality.

We use $\SigmaPS$ in 
order to monitor the size of perturbative discretisation effects, making use of
lattice perturbation theory to $O(g_0^2)$; clearly this is meaningful only
at high energy (SF) scales. 
We first define the ratio of the lattice SSF $\SigmaP(u, a/L)$,
computed at 1-loop, to the same quantity in the $L/a=\infty$ limit, in order to determine
the numerical effect of all lattice artefacts appearing at this order:
\begin{gather}
\label{eq:R_P_PT}
\frac{\SigmaP^{\rm\scriptscriptstyle 1-loop}(u, a/L)}{\sigmaP^{\rm\scriptscriptstyle 1-loop}(u)}
\,\, =  \,\, 1 + u \delta_{\rm\scriptscriptstyle{P}}(a/L) \,\,, \\
\delta_{\rm\scriptscriptstyle{P}}(a/L) = -d_0 \ln(2) c_{\rm\scriptscriptstyle{P}}(a/L) \,\, .
\label{eq:delta}
\end{gather}
In the above $d_0= 8/(4\pi)^2$ is the universal anomalous dimension coefficient for the pseudoscalar bilinear 
operator. Analogous expressions are defined for the scalar operator (same $d_0$).
The numerical values of $c_{S,P}(a/L)$, calculated in ref.~\cite{chiSF-PT-TEMP}, are given in Table.~\ref{tab:cSP}. 
\begin{table}[th]
\begin{tabular}{c | c c}
 $L/a$ & $c_{\rm\scriptscriptstyle S}(a/L)$ & $c_{\rm\scriptscriptstyle P}(a/L)$ \\
\hline 
 6 & 0.1080 & 0.0486\\
 8 & 0.0688 & 0.0458\\
 12 & 0.0240 & 0.0168\\
 16 & 0.0121 & 0.0086\\
\end{tabular}
\caption{Subtraction coefficients used to remove discretisation effects
from the nonperturbative step-scaling functions $\Sigma_{\rm\scriptscriptstyle{S,P}}$ up to
$O(g^2)$ as given in Eqs.~\eqref{eq:R_P_PT}, \eqref{eq:delta}.
\label{tab:cSP}}
\end{table}

The lattice artefacts of $O(g_0^2 a^n)$ are subsequently subtracted from the $\Sigma_{\rm\scriptscriptstyle{S,P}}$ 
functions, computed nonperturbatively as in \req{eq:Ssf-PchiSF} and \req{eq:Ssf-SchiSF},
according to
\begin{gather}
\label{eq:Sigma_sub}
\Sigma^{\text{sub}}_{\rm\scriptscriptstyle{S,P}}(u, a/L) \equiv 
\frac{\Sigma_{\rm\scriptscriptstyle{S,P}}(u, a/L)}{1 + u \delta_{\rm\scriptscriptstyle{S,P}}(a/L)} \,\, .
\end{gather}
The remaining discretisation errors in $\Sigma^{\text{sub}}_{\rm\scriptscriptstyle{S,P}}$ are $O(g_0^4 a^2)$. 

Supressing all $O(g_0^2a^n)$ terms using lattice perturbation theory,
we can in principle remove the corresponding parameters in our global fit ansatze; cf.\ Eq.~(\ref{Sigma_S/P}).
This means that we can expect more accurate determination of the remaining
parameters, and more robust determinations e.g.\ upon increasing
the order at which the expansion is truncated.
Furthermore we might expect, especially at high energies where the gauge coupling is
small, that removing lattice artefacts up to this order removes the
largest contribution at each fixed order in $a^n$.
This means that we may expect smaller coefficients in the remaining power series after the
subtraction (i.e.\ the resulting power series is better behaved), 
although this is not guaranteed. Thus we would expect our fit forms, 
which are truncated to some order, to
represent the subtracted data more accurately,
resulting in better $\chi^2/\text{dof}$ and improved confidence 
in our extrapolated results
\begin{equation} \label{Sigma_S/P}
\SigmaPS(u, a/L) =
1 + \sum_{i,j=1} d_{ij} u^i \Bigl(\frac{a}{L}\Bigr)^{2j} \,\, ,
\end{equation}
where terms depending on $\ln(a/L)$ are again neglected.

\begin{figure}[ht]
\includegraphics[width=0.5\textwidth]{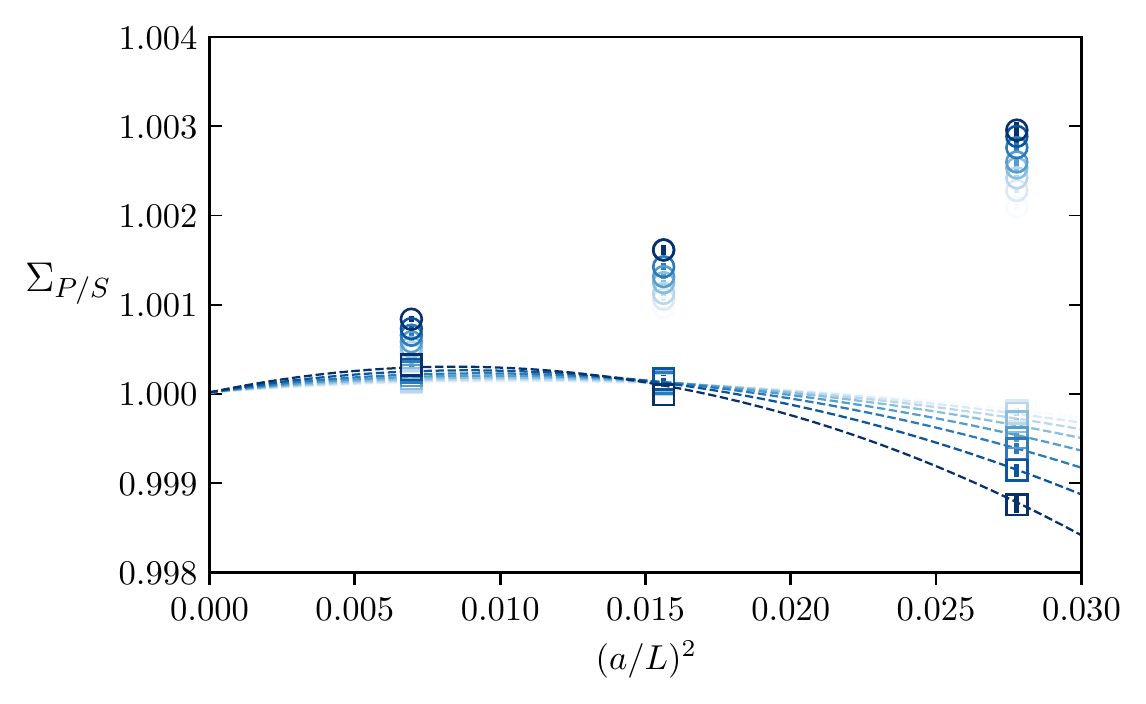}
\caption{Global fit of $\Sigma_{{\rm\scriptscriptstyle P}/{\rm\scriptscriptstyle S}}$ data in the high energy (SF) regime
after subtraction of $O(u, (a/L)^n)$-effects. Data before/after subtraction
are given by open circles/squares. Fit results to the subtracted data are
shown as dashed lines.
\label{fig:SP-global-sub}}
\end{figure}

Here we test these expectations for the ratio of step-scaling functions
$\Sigma_{{\rm\scriptscriptstyle P}/{\rm\scriptscriptstyle S}}$. Since the scalar and pseudoscalar
bilinear operators have the same continuum anomalous dimension, the deviation
of this quantity from one is a measure of lattice artefacts. For a fixed
renormalised coupling $u$, the data should extrapolate to 1 in the
$a/L \to 0$ limit. Fig.~\ref{fig:SP-global-sub} shows
$\Sigma_{{\rm\scriptscriptstyle P}/{\rm\scriptscriptstyle S}}$ vs.\ $(a/L)^2$ for the eight
renormalised couplings in the high-energy
(SF) regime, both before and after the subtraction specified by Eq.~(\ref{eq:Sigma_sub}).
The strong statistical correlation of $\SigmaP$ and $\SigmaS$ on a given ensemble
results to extremely small statistical uncertainty, 
leaving systematic effects to dominate.
The raw data is shown above (circles, w/out lines passing through them), with the darker colours corresponding
to larger renormalised coupling. We observe that the data at smaller coupling
is nearer to 1, as is expected for a ``well-behaved'' series at small coupling
and sufficiently small lattice spacing. 
A fit of the data to Eq.~\eqref{Sigma_S/P} 
for $(i, j) = (3, 2)$ gives a $\chi^2/{\rm d.o.f.} \sim 0.48$. Increasing
$i_{\text{max}}$ or $j_{\text{max}}$ in the fit doesn't appreciably improve
the $\chi^2/{\rm d.o.f.}$.

The lower part of Fig.~\ref{fig:SP-global-sub} gives the same data 
but after subtraction specified by~\eqref{eq:Sigma_sub}, along
with the curves determined by a global fit to this data. 
For $L/a=8$ and 12,
the subtracted data is exceptionally close to 1, indicating that the
$O(g_0^2)$ lattice artefacts are indeed dominant. There is a more obvious
discrepancy from 1 for $L/a=6$, up to around 0.001 for the $u=2.0120$ ensemble,
but the overall size of this is 
significantly smaller than for
the unsubtracted data, indicating the leading source of artefacts are still
removed. Note that after subtraction the data at smaller coupling is still
closer to one than the data at larger coupling.
We can fit the subtracted data to the form Eq.~\eqref{Sigma_S/P},
but now excluding the terms $d_{1j}$ that should be absent from the 
subtraction. For the same $(i, j) = (3, 2)$ fit as in the unsubtracted case,
we find a $\chi^2/{\rm d.o.f.}$ of 0.96. This is shown as the dotted lines
in Fig.~\ref{fig:SP-global-sub}. As expected, we note that
increasing $(i_{\text{max}}, j_{\text{max}})$ improves the 
$\chi^2/{\rm d.o.f.}$ somewhat and that fits to the unsubtracted
data using the restricted form $d_{1j}=0$ result in poor $\chi^2$.

\begin{figure}[ht]
\includegraphics[width=0.5\textwidth]{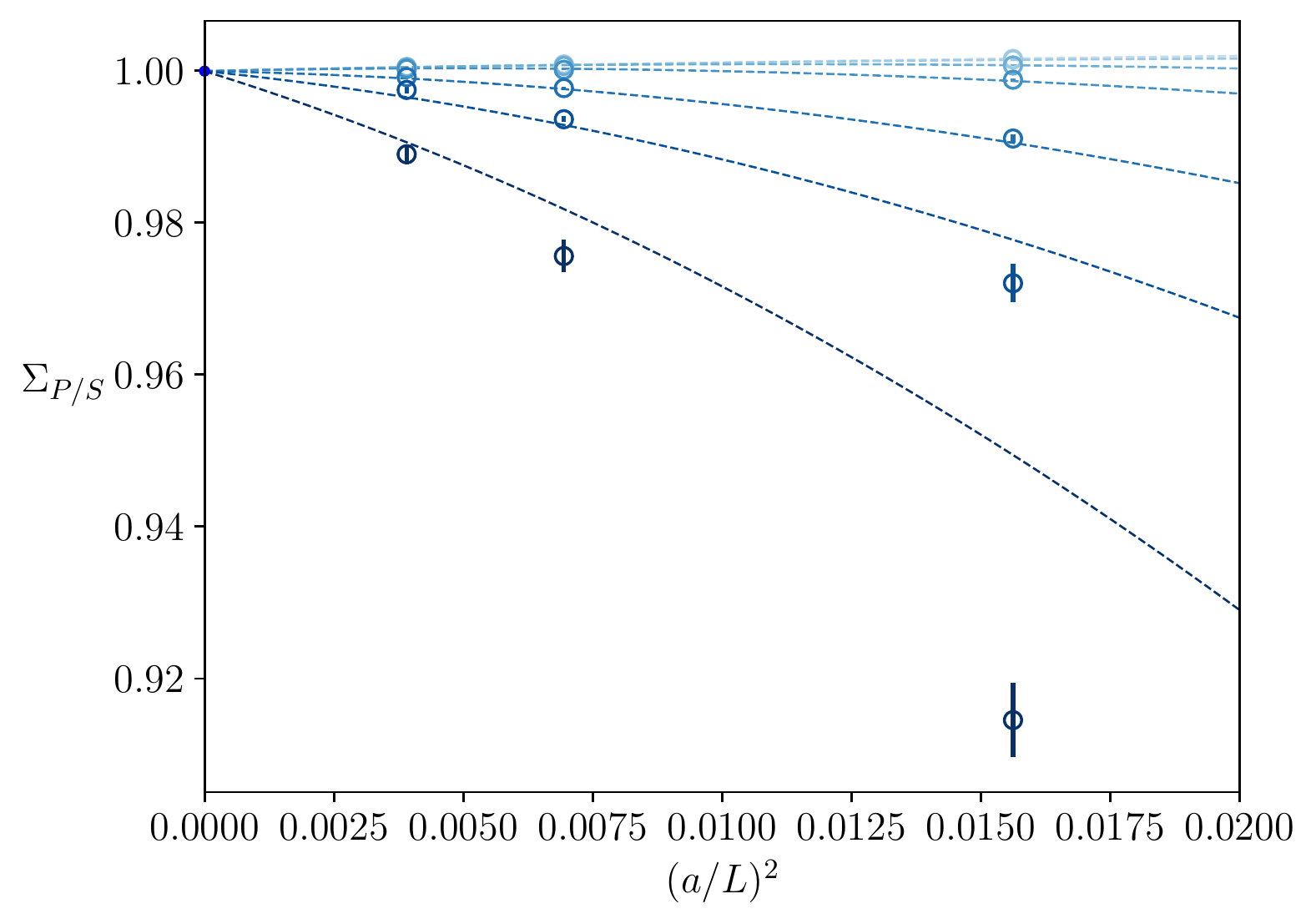}
\caption{Global fit of $\Sigma_{{\rm\scriptscriptstyle P}/{\rm\scriptscriptstyle S}}$ data in the low energy (GF) regime.
\label{fig:SP-gf}}
\end{figure}

The analogous study of the ratio $\SigmaP/\SigmaS$ 
in the low energy regime, shown in Fig.~\ref{fig:SP-gf},
does not involve any perturbative subtractions. 
The results reveal that higher order powers
of $(a/L)^2$ are present, and become increasingly pronounced, at larger coupling.
In fact, at our two largest couplings ($u=5.8673$ and $u=6.5489$), 
a low-order polynomial
in $(a/L)^2$ has trouble capturing the behaviour,
indicating that in $\SigmaS$ or $\SigmaP$ (or both) an extrapolation to the
continuum value may be unreliable for the given lattice extents and couplings.
The result of fitting the data to Eq.~\eqref{Sigma_S/P} for
$(i_{\text{max}}, j_{\text{max}}) = (4, 3)$ is shown 
in Fig.~\ref{fig:SP-gf}, which returns a $\chi^2/{\rm d.o.f.}$ of 4.2.
Increasing $(i_{\text{max}}, j_{\text{max}})$ improves the
$\chi^2/{\rm d.o.f.}$ somewhat, but all fits studied have trouble with
the $L/a=8$ points for large coupling. 
On the other hand the $(4, 3)$ fit has
$\chi^2/{\rm d.o.f.} = 1.3/0.66$ if data from the last/second to last couplings 
are removed.

\section{Quark mass running at high energies}
\label{sec:results-he}
We now turn to the computation of the step-scaling functions themselves,
which are the main inputs for the determination of the nonperturbative
running of the quark masses. On each pair of ($L$, $2L$) ensembles
we compute $\SigmaPchiSF$, defined in Eq.~(\ref{eq:Ssf-PchiSF}); henceforth 
the superscript $\chi$SF will be dropped:
\begin{equation}
\SigmaP(u, a/L) = \frac{\ZP(u, 2L/a)}{\ZP(u, L/a)} \,.
\end{equation}
In both the high energy and low energy regimes we work at three
different lattice spacings, except for the largest coupling in the high energy
range (the switching point, $u=2.0120$) where we use four lattice spacings. 
We note here
that the values of $\SigmaP$ at different couplings 
are statistically uncorrelated. Furthermore the
numerator and denominator are uncorrelated and as a result the statistical
uncertainties are significantly larger than for the quantities 
$\SigmaPS$ or $\ZS/\ZP$ studied in Sec.~\ref{sec:zratio}. 
In order to effectively leverage the data in the high-energy (SF) regime,
we carry out a global to the following form:
\begin{equation} \label{sigmaP_fit}
\SigmaP(u, a/L) = 1 + \sum_{i=1, j=0} \, b_{ij} u^i (a/L)^{2j} \, .
\end{equation}
The continuum step-scaling function $\sigmaP$ is then given by
$\sigmaP(u) = \lim_{a \to 0} \SigmaP(u, a/L)$. Although 
$\SigmaP(g_0^2, a/L)$ is computed at specific values of the
bare coupling and lattice volume, we are interested in its behaviour
with varying renormalised coupling; hence the notation $\SigmaP(u, a/L)$.

The continuum coefficient $b_{10} = -d_0 \ln(2)$ is known from 
perturbation theory, where $d_0 = 8/(4\pi)^2$ is the universal lowest-order 
quark mass anomalous dimension; see Eqs.~(\ref{eq:tau-beta-PT}) below.
Also known in perturbation theory is the 
coefficient $b_{20} = -d_1 \ln(2)+(d_0^2/2 - b_0 d_0)\ln(2)^2$. Here 
$b_0 = (11 - 2\NF /3)/(4\pi)^2$ is the universal lowest-order coefficient of the 
Callan-Symanzik $\beta$-function and $d_1 = 1/(4\pi)^2 (0.2168 + 0.084\NF)$, 
the NLO coefficient of the quark mass anomalous dimension, is specific to the SF
 scheme and was computed in \cite{Sint:1998iq}. For $\NF = 3$ we have 
 $b_{20} = -0.0028$. 
We can constrain the fit form Eq.~\eqref{sigmaP_fit} using these values.
By subtracting leading $O(u)$ discretisation errors from our data 
(cf.\ Eq.~(\ref{eq:Sigma_sub})), we
can also set the terms $b_{1j>0}$ to zero. However we find
that doing so generally leads to somewhat higher $\chi^2/\text{dof}$
values (and is less conservative, giving smaller errors), 
and so we choose to work with the unsubtracted data.

For the fits considered here we take 
$(i_{\text{max}}, j_{\text{max}}) = (3, 2)$.
When fitting the full dataset, we find $\chi^2/\text{dof} \approx 1.4$,
and this is not appreciably improved by increasing 
$(i_{\text{max}}, j_{\text{max}})$. 
We attribute this to a partial breakdown of the polynomial
form Eq.~\eqref{sigmaP_fit} when including the $L/a = 6$ points in
our dataset, especially at large couplings. If we remove
these points we find an improved $\chi^2/\text{dof} =0.95$.
If we leave the term $b_{20}$ unconstrained this fit returns
a value of $b_{20} = -0.0019(11)$, compatible with the perturbative
value. Including the $L/a = 6$ data instead gives $b_{20} = -0.0012(9)$,
once again giving some evidence that the form~\eqref{sigmaP_fit}
strains to represent these points. Therefore for our preferred fit
we drop the $L/a=6$ data and fix $b_{20}$ to its perturbative value.

\begin{figure}
\includegraphics[width=0.5\textwidth]{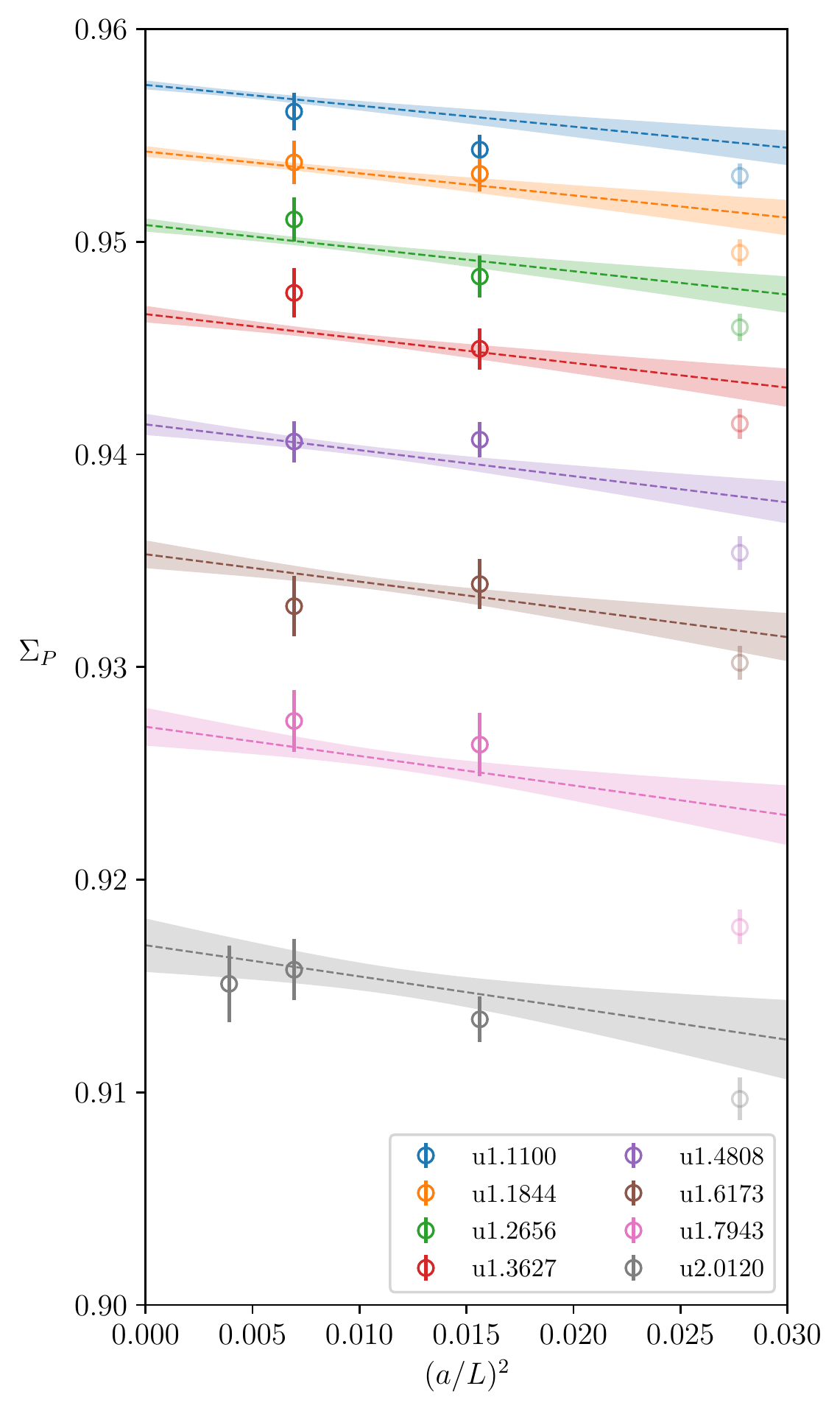}
\caption{Results of a global 
fit to the step-scaling data $\SigmaP(u, a/L)$ in the high-energy
regime. The open circles give the raw data while the filled bands are
the results returned from the fit at the respective $u$ values.
The transparent $L/a = 6$ data points are not included in the fit.
The data points and the bands of the same colour are at a fixed value 
of the renormalised squared coupling $u$.
\label{fig:sigmaP-global}}
\end{figure}

The raw step-scaling data in the high energy regime, along with the 
curves from the global
fits evaluated at their
respective $u$ values, are shown in Fig.~\ref{fig:sigmaP-global}.

\begin{figure}
\includegraphics[width=0.5\textwidth]{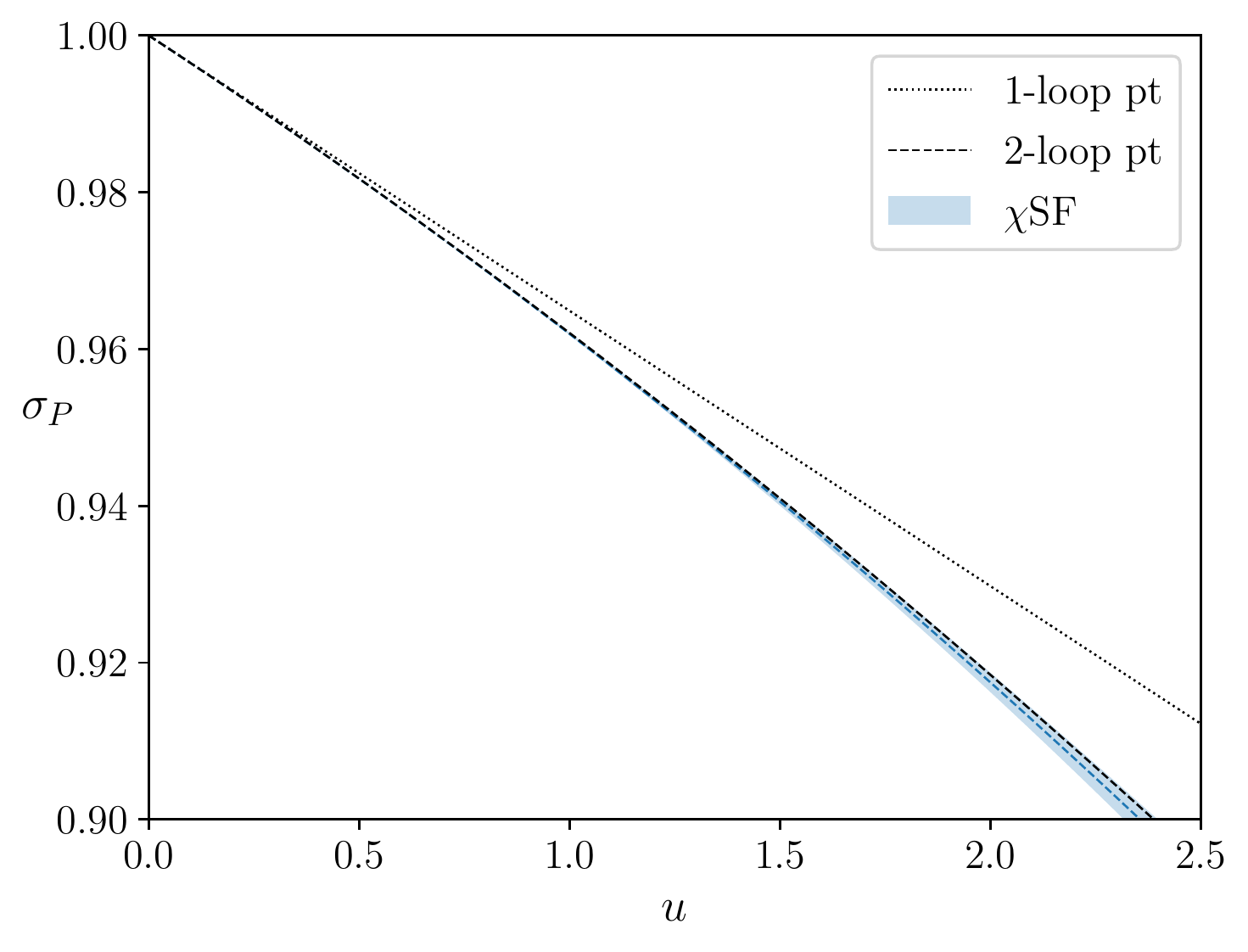}
\caption{Continuum limit of the nonperturbatively determined step-scaling
function $\sigmaP(u)$, compared with perturbation theory. 
The 1-loop perturbative result
is universal while the 2-loop result is specific to the $(\chi)$SF 
renormalisation scheme.
\label{fig:sigmaP_vs_u}}
\end{figure}

The continuum curve $\sigmaP(u)$ obtained from the fit (\ref{sigmaP_fit})
is shown in Fig.~\ref{fig:sigmaP_vs_u} 
and compared with the expectations from perturbation theory. One sees
that in the high energy region the nonperturbative result agrees 
well with the two-loop result from perturbation theory, indicating
that the perturbative series is fairly well converged. This result is 
also consistent with the findings of~\cite{Campos:2018ahf}, 
and demonstrates the universality of $\chi$SF and SF.

We have established that the result shown in 
Fig.~\ref{fig:sigmaP_vs_u} is a robust nonperturbative
estimate of $\sigmaP(u)$. Having been constrained
by perturbation theory at small couplings,
it is also valid below the lowest simulated 
value $\uSF \sim 1.1$. In other words, this result 
can be used in the whole energy range 
above $\mu_0/2 \sim 2~$GeV, allowing us to compute
the mass-evolution values $R^{(k)}$, defined as (cf.\ Eq.(\ref{sigmaP-mbar}))
\begin{equation} \label{Rk}
R^{(k)} = \frac{\mbar(2^k \mu_0)}{\mbar(\mu_0/2)} = 
\prod_{n=0}^{k} \sigmaP(u_n) \,\, ,
\end{equation}
for arbitrarily large energy scales $2^k \mu_0$ (large $k$);
in the above $u_n = \uSF(2^n \mu_0)$.
For small values of $k$, Table~\ref{tab:Rk} shows our 
results for $R^{(k)}$ from the fit 
to the full data set (labelled $R^{(k)}$) and those excluding the $L/a=6$ 
data points (labelled $R^{(k)}$-w/o $6$). The results
including the $L/a=6$ data are systematically larger than those without,
an effect consistent with the findings of~\cite{Campos:2018ahf}. As 
previously stated, and in accordance with ref.~\cite{Campos:2018ahf},
we take our preferred values to be those excluding $L/a=6$ since there is 
evidence that higher order discretisation effects are large for these points, 
and the uncertainty estimate is more conservative when excluding them.
Having excluded the $L/a=6$ data from our analysis, we compute $R^{(k)}$
for increasing $k$ values, which take us beyond the energy range covered by
our data. This is tenable, given that our fit is also constrained by perturbation 
theory at small $u$; cf.\ Fig.~\ref{fig:sigmaP_vs_u}. The behaviour of $R^{(k)}$
 with growing $k$ is displayed in Table~\ref{tab:Rk-M/m}, where we also
show results for $M/\mbar(\mu_0/2)$ (see Eq.\eqref{eq:M/bar-m} below).
We observe that $M/\mbar(\mu_0/2)$ remains constant within its error as
$k$ increases.

\begin{table}[th]
\begin{tabular}{c c c c c c}
$k$ & $u_k$ & $R^{(k)}$-w/o 6 & $R^{(k)}$ & $R^{(k)}$~\cite{Campos:2018ahf} \\
\hline
0 & 2.0120     &0.9169(13)  & 0.9191(10) & 0.9165(12)  \\
1 & 1.7126(31) & 0.8536(19) & 0.8569(16) & 0.8530(17)  \\
2 & 1.4939(38) & 0.8031(22) & 0.8070(17) & 0.8025(20) \\
3 & 1.3264(38) & 0.7615(24) & 0.7656(19) & 0.7608(21) \\
4 & 1.1936(35) & 0.7263(25) & 0.7307(20) & 0.7257(22) \\
5 & 1.0856(32) & 0.6961(25) & 0.7005(20) & 0.6968(24) 
\end{tabular}
\caption{Mass ratios $R^{(k)}$, obtained from nonperturbative SSF in $\chi$SF for increasing $k$-values,
compared with the preferred SF fit from ref.~\cite{Campos:2018ahf}. The column marked $R^{(k)}$-w/o 6 gives results
excluding the $L/a=6$ data points in the fit.
\label{tab:Rk}}
\end{table}

\begin{table}[th]
\begin{tabular}{c c c c}
$k$ & $u_k$ & $R^{(k)}$-w/o 6 & $M/\mbar(\mu_0/2)$\\
\hline
  5 & 1.0856(32) & 0.6961(25) & 1.7514(63) \\
10 & 0.7503(19) & 0.5899(25) & 1.7519(74)  \\
20 & 0.4664(8) & 0.4769(22) & 1.7523(81) \\
30 & 0.3392(4) & 0.4136(19) & 1.7522(80)\\
40 & 0.2667(3) & 0.3716(17) & 1.7524(80)
\end{tabular}
\caption{Mass ratios $R^{(k)}$ and the ratio of 
RGI to renormalised quark mass at scale $\mu_0/2$,
obtained from nonperturbative SSF in $\chi$SF
at large $k$-values.}
\label{tab:Rk-M/m}
\end{table}

Finally, we can construct the running factor that takes a renormalised
quark mass in our chosen ($\chi$)SF scheme at the scale $\mu_0/2$ to the
renormalisation group invariant quark mass $M$:
\begin{equation}
\label{eq:M/bar-m}
\frac{M}{\mbar(\mu_0/2)} = \Big[ \frac{M}{\mbar(2^k \mu_0)} \Big ]\, \, \Big[\frac{\mbar(2^k \mu_0)}{\mbar(\mu_0/2)} \Big ] \, .
\end{equation}
The first factor on the r.h.s. can be calculated from 
\begin{multline}
\frac{M}{\mbar(2^k \mu_0)} = 
\label{eq:M-barm-2k}
[2 b_0 \gbar^2_{\SF}(2^k \mu_0)]^{-d_0/2 b_0} \times \\
\exp \Bigl\{ -\int_0^{\gbar_{\SF}(2^k \mu_0)} dx 
\Bigl[\frac{\tau(x)}{\beta(x)} - \frac{d_0}{b_0 x} \Bigr]
\Bigr\} \,,
\end{multline}
with $\tau$ and $\beta$ given by their perturbative expressions
\begin{eqnarray}
\label{eq:tau-beta-PT}
\tau(x) &=& -x^2[d_0 + d_1 x^2 + d_2 x^4 + \cdots] \, ,\\
\beta(x) &=& -x^3[b_0 + b_1 x^2 + b_2 x^4 + b_3 x^6 + \cdots] \, .\nonumber
\end{eqnarray}
Specifically, we use the 2-loop result for $\tau(x)$ (i.e.\ $d_0$ and $d_1$~\cite{Sint:1998iq})
and the 3-loop result for $\beta(x)$ (i.e.\ $b_0$, $b_1$, and $b_2$~\cite{Caswell:1974gg,Jones:1974mm,Bode:1999sm}), supplemented by an estimate of $b_3$ from a fit performed in~\cite{DallaBrida:2018rfy}).
The second factor is $R^{(k)}$ of Eq.~(\ref{Rk}) and it is known nonperturbatively.
Having previously shown that the result and its error are practically
independent of $k$, we quote for $k =10$:
\begin{align}
\label{eq:M-barm-SF}
\frac{M}{\mbar(\mu_0/2)} = 1.7519(74) \, .
\end{align}

The above result agrees with the value obtained in the SF-LQCD setup of ref.~\cite{Campos:2018ahf},
$M/\mbar(\mu_0/2) = 1.7505(89)$. This is the outcome of a detailed analysis performed by the authors,
consisting of four different extrapolation procedures of the $\SigmaP(u, a/L)$ data, from which
estimates of $M/\mbar(\mu_0/2)$ are extracted. Their preferred result, quoted here, is obtained
from their so-called $\tau$:\emph{global}--\texttt{FITB*} procedure, which consists in performing the
continuum extrapolation of $\SigmaP(u,a/L)$ and the determination of the anomalous dimension 
$\tau(\gbar)$ simultaneously. We have also applied this procedure to our data. Outlining the method,
we start by rewriting Eq.(\ref{sigmaP_fit}) as:
\begin{equation} \label{sigmaP_fit-trunc}
\SigmaP(u, a/L) = \sigmaP(u) +  \, \sum_{n=0}^2 \rho_n u^n (a/L)^{2} \, .
\end{equation}
Note that have been dropped from Eq.(\ref{sigmaP_fit}) terms of $O(a/L)^4$ and higher (i.e.\ terms with $j \ge 2$).
We have also dropped terms  multiplying $(a/L)^2$ of $O(u^3)$ and higher (i.e.\ $b_{n1}$ terms with $n \ge 3$).

We write the logarithm of $\sigmaP(u)$, in terms of the anomalous dimension $\tau$, as (cf.\ Eq.~(\ref{sigma_P}))
\begin{equation}
  \ln\left(\SigmaP(u,a/L) - \sum_{n=0}^2 \rho_n u^n (a/L)^{2} \right) = 
  - \int_{\sqrt{u}}^{\sqrt{\sigma(u)}} \dif x\, \frac{\tau(x)}{\beta(x)}\,,
\end{equation} 
where $\sigma(u)$ is the step scaling function of the renormalised gauge coupling; i.e.\
for $\gbar^2(\mu)=u$, $\sigma(u) = \gbar^2(\mu/2)$.  For the integrand on the r.h.s. we use
the truncated expansions of Eqs.~(\ref{eq:tau-beta-PT}), where $d_0, d_1, b_0, b_1, b_2$ are
taken from perturbation theory and $b_3$ from a fit as explained above. Finally, a global fit of the 
$\SigmaP(u,a/L)$ data, with free fit parameters $\rho_0, \rho_1, \rho_2$ and $d_2$,
results to a continuum expression for $\tau(u)$.

Having obtained $\tau(u)$, we can now work out directly $M/\mbar(\mu_0/2)$, using Eq.(\ref{eq:M-barm-2k}),
with the scale $\mu_0/2$ in place of $2^k \mu_0$. This gives 
\begin{align}
\label{eq:M-barm-SF2}
\frac{M}{\mbar(\mu_0/2)} = 1.7517(81) \, .
\end{align}
There is excellent agreement with the result of the first procedure, cf.\ Eq.(\ref{eq:M-barm-SF}), as well as with the SF-LQCD result of ref.~\cite{Campos:2018ahf}. We find this particularly encouraging, given the different philosophy of the two procedures.
The first one entails a choice of $k$ in Eqs.(\ref{Rk}) and (\ref{eq:M/bar-m}), which we have taken to be $k=10$. We have checked the stability of our results for $5 \lesssim k \lesssim 40$. The parametrisation of discretisation effects included $(a/L)^4$ contributions. In the second procedure, these were taken to be $(a/L)^2$. Although we could have included $(a/L)^4$ terms, we opted to stay as close as possible to the choices made in ref.~\cite{Campos:2018ahf}.

\section{Quark mass running at low energies}
\label{sec:results-le}
Having computed the running factor to convert the renormalised mass at
the scale $\mu_0/2$ to the renormalisation group invariant mass, we now
turn to the computation of the running factor in the low-energy (GF) regime.

\begin{figure}
\includegraphics[width=0.5\textwidth]{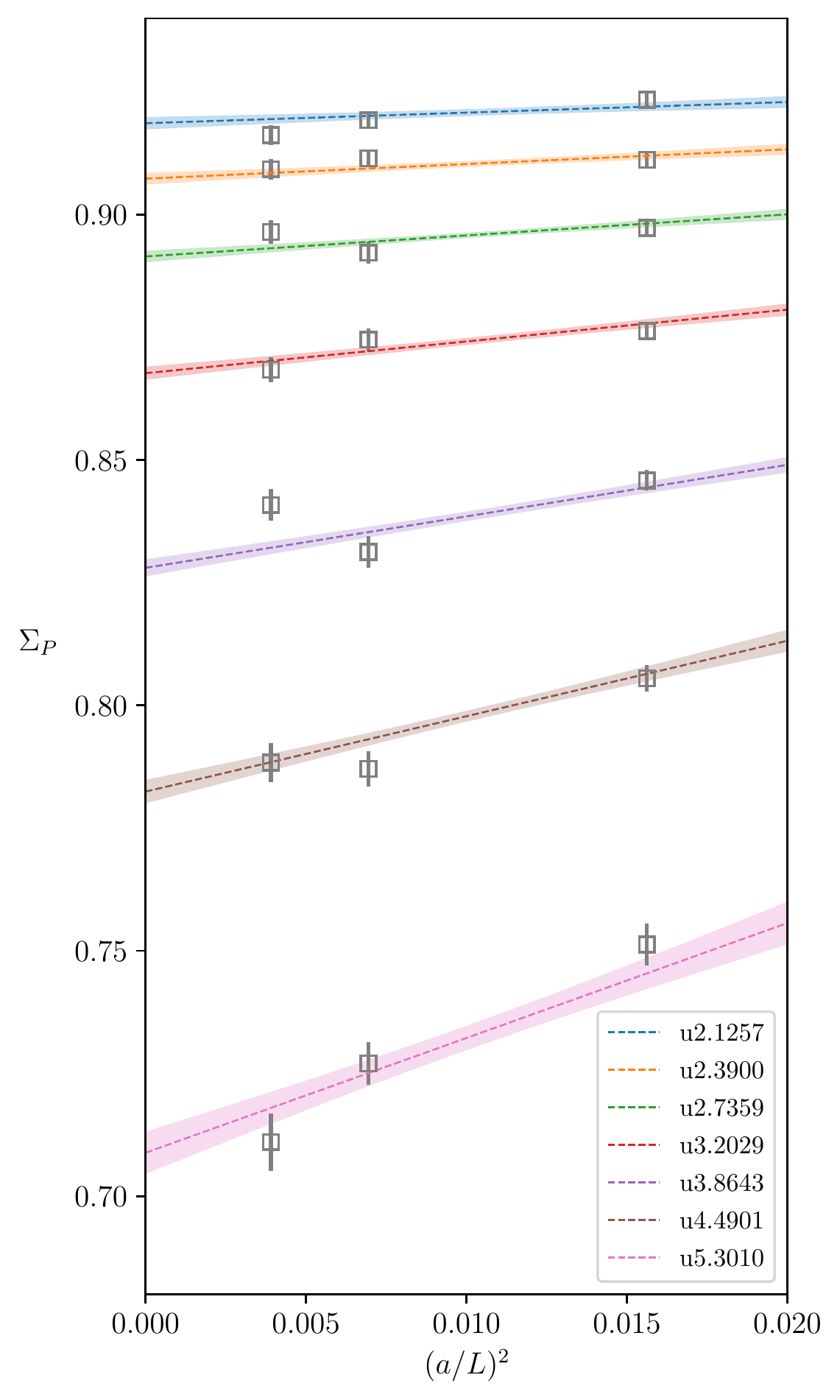}
\caption{Results of a global 
fit to the step-scale data $\SigmaP$ in the low-energy
regime. The open squares give the raw data while the filled bands are
the results returned from the fit at the respective $u$ values.
The data points close to the bands are at {\it approximately} the same $u$.
\label{fig:sigmaP-gf}}
\end{figure}

Whereas in the high-energy (SF) regime, we had lattices of extent
$L/a=6,8,12$ (and $L/a=16$ at $u=2.0120$), in the GF regime our lattices have
extent $L/a=8,12,16$, which should improve the continuum extrapolation.
In the high energy regime we found that discretisation effects increase as
the coupling is increased, and we observe a similar trend in our data in the
low energy regime. It is evident from Fig.~\ref{fig:sigmaP-gf} that the $(a/L)^2$ 
coefficient grows with increasing coupling.

The low energy hadronic scale $\mu_{\text{had}}$ is defined by
\begin{equation}
u(\mu_{\text{had}}) = 9.25 \,,
\end{equation}
corresponding to a physical scale 
$\mu_{\text{had}} = 233(8)$~MeV~\cite{Bruno:2017gxd}. Since the ratio of the
switching scale $\mu_0/2 \sim 2$~GeV to the hadronic scale $\mu_{\text{had}}$ is 
not a power of two, it is inconvenient to carry out the analysis in terms of the step
scaling function $\sigmaP(u)$, which only expresses the quark mass running between
consecutive scales $\mu$ and $\mu/2$. It is preferable to rely on the mass anomalous
dimension $\tau(g)$, related to $\sigmaP(u)$ through
\begin{equation} \label{sigma_P}
\sigmaP(u) = \exp{\Bigl[
-\int_{\sqrt{u}}^{\sqrt{\sigma(u)}} dg \frac{\tau(g)}{\beta(g)}
\Bigr]} \,.
\end{equation}
Expanding the integrand of the above equation $ f(g) \equiv \tau(g)/\beta(g)$ as a
power series
\begin{equation}
f(g) = \frac{1}{g}\sum_{k=0} f_{k} \, g^{2k} \,\,,
\end{equation}
we obtain the following expression for the step scaling function:
\begin{equation} \label{lnsigmaP}
\ln \sigmaP (u) = - \sum_{k=0} f_{k} \int_{\sqrt{u}}^{\sqrt{\sigma(u)}} dg \,\, g^{2k-1}
 \,\,.
\end{equation}
Unlike what we did in the high-energy regime, no input from perturbation theory is introduced in
the above expression.  

The lattice step scaling function $\SigmaP(u,a/L)$ is given as a series expansion in Eq.~(\ref{sigmaP_fit}).
This can be conveniently rearranged as
\begin{equation} \label{SigmaP_fit_le}
\SigmaP(u, a/L) = \sigmaP(u) \Big [ 1 +  \sum_{j=1} (a/L)^{2j} \sum_{i=1} d_{ji} u^i \Big ] \,\, ,
\end{equation}
resulting to the fit function
\begin{equation} \label{lnSigmaP_fit_le}
\ln [\SigmaP(u, a/L)] = \ln [\sigmaP(u) ]  + \Big [ \sum_{j=1} (a/L)^{2j} \sum_{i=1} d_{ji} u^i \Big ] \,\, .
\end{equation}
Thus the data points $\ln \SigmaP$ on the l.h.s. are to be fit by the product of the series of $\ln \sigmaP$
of Eq.(\ref{lnsigmaP}) times the double series with coefficients $d_{ji}$.

We can vary the number of coefficients $f_k$ parameterising
the continuum form up to a value $k_{\text{max}}$. We also vary
the number of coefficients $d_{ji}$ that parameterise lattice
artefacts as follows: We consider artefacts that scale as $(a/L)^2$
as well as $(a/L)^4$; i.e.\ we set $j_{\text{max}} = 2 $.
For each of the allowed values $j = 1,2$, we vary the order
of the polynomial in $u$ up to $i_{\text{max}}(j)$.
Our preferred fit is the one with $k_{\text{max}}=4$, $i_{\text{max}}(1) = 4$,
$i_{\text{max}}(2) = 0$. The fit has a $\chi^2/\text{dof}$ of 1.1.

\begin{figure}[ht]
\includegraphics[width=0.5\textwidth]{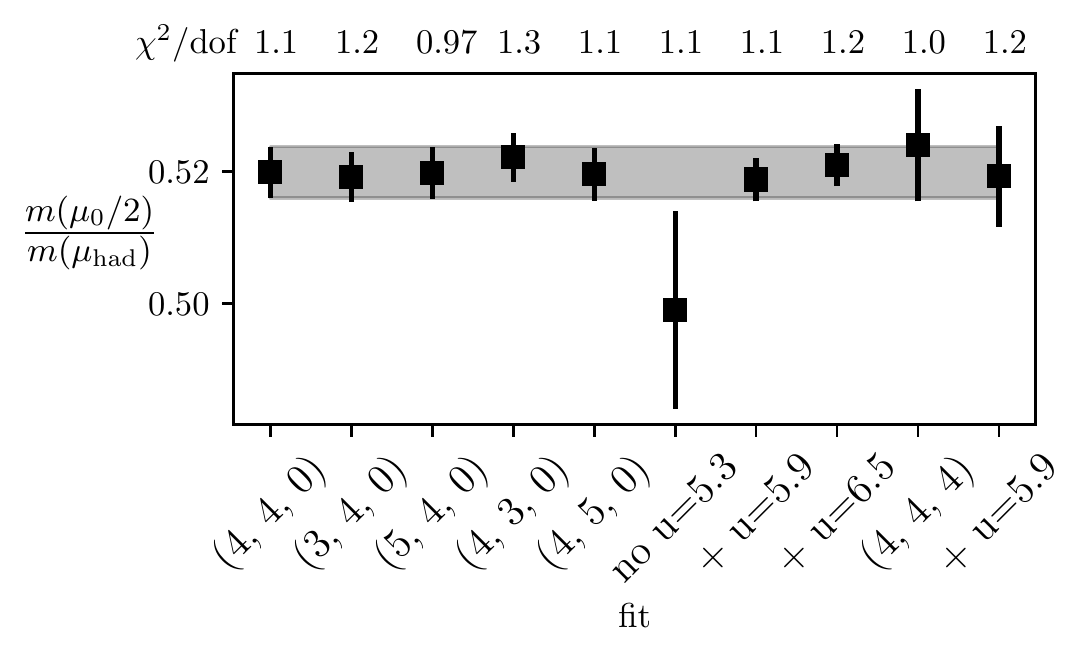}
\caption{Stability analysis of the mass running factor in the low-energy
regime. The leftmost point shows the result
of our preferred fit~(Eq.~\eqref{gf_res}). The indices 
$(k_{\text{max}}, i_{\text{max}}(1), i_{\text{max}}(2))$
on the x-axis give the number of
polynomial terms in $u$ used to parameterise the continuum, 
$(a/L)^2$, and $(a/L)^4$ dependences respectively.
Moving from left to right,
the labels `no $u=5.3$', `$+ u=5.9$', etc.\ show the effect of removing/adding
data at the specified couplings to our preferred (4, 4, 0) fit.
The rightmost point shows the effect of adding $u=5.9$ data to the (4, 4, 4)
fit.
\label{fig:stability}}
\end{figure}

We have checked the stability of our final result 
to changes in parameters controlling
the fit form. We find good agreement in the result using different fit forms,
with $\chi^2/\text{dof} \approx 1$, indicating our data is well-represented
by and relatively insensitive to the precise details of the fit.
This is shown in 
Fig.~\ref{fig:stability} for fit forms labelled
$(k_{\text{max}}, i_{\text{max}}(1), i_{\text{max}}(2))$. In addition, we can also
include/exclude the data at different couplings used in the fits.
The authors of~\cite{Campos:2018ahf} excluded data at the two highest couplings 
$u=5.8673$ and $u=6.5489$. Including these couplings gives a consistent
result but with somewhat smaller errors. If one instead removes the
next high coupling $u=5.3010$, the data is not sufficiently constraining
in the low energy region and the error increases significantly.
The results from these variations are shown using our preferred (4, 4, 0) 
(and (4, 4, 4)) parameterisation in Fig.~\ref{fig:stability}.

After having fit our data for $\sigmaP$ in both the SF and GF regimes
and taken the continuum limit, we find at the switching scale
\begin{align}
\sigma_{\rm\scriptscriptstyle P,SF}(\mu_0/2) &= 0.8951(23) , \\
\sigma_{\rm\scriptscriptstyle P,GF}(\mu_0/2) &= 0.8941(12) \,. 
\end{align}
The compatibility of these results at the threshold scale $\mu_0/2$,
where the definition of the renormalised coupling changes, is yet another
indicator of the robustness of our analysis.

Having obtained $f(g)$ from the fit and using
the polynomial expression for $\beta(g)$ given in~\cite{Campos:2018ahf}, 
we can reconstruct the function $\tau$ and determine 
\begin{equation} \label{gf_res}
\frac{\mbar(\mu_0/2)}{\mbar(\mu_{\text{had}})} = 0.5199(39) \, ,
\end{equation}
which can be compared with the result from~\cite{Campos:2018ahf},
$\frac{\mbar(\mu_0/2)}{\mbar(\mu_{\text{had}})} = 0.5226(43)$.

\section{Conclusion}
\label{sec:concl}
We have analysed the ensembles generated for the computation of
the RG-running of the quark mass in $N_f=3$ massless QCD
(with SF boundaries)~\cite{Campos:2018ahf},
imposing chirally rotated Schr\"odinger functional boundary conditions
on the valence quarks.
The data spans a few orders of magnitude, allowing a completely 
nonperturbative determination of the mass running function between the hadronic 
and very high energy scales, where contact with fixed-order perturbation theory 
can be made. Our computations are characterised by two different definitions of the renormalised 
gauge coupling below and above an energy threshold (switching scale) of $\sim 2$~GeV. This
results to some differences in the computational strategies in the low- and high-energy
regimes.

In order to recover all the symmetries of QCD in the continuum limit,
we have performed the required
nonperturbative tuning  of the boundary counterterm $\zf$ of the $\chi$SF valence quark.
The critical value $\hopc$ of the mass tuning parameter $\kappa$ is taken from ref.~\cite{Campos:2018ahf}.

We computed in both high- and low-energy regions the ratio $\ZS/\ZP$ and the ratio of
step-scaling functions $\SigmaS/\SigmaP$. For the former we find that
results match smoothly with 1-loop PT in
the high energy regime. We also find consistency with determinations based on
Ward identities in the low energy regime. The latter provides an important diagnostic 
check for the validity of our fit forms, and we find that the one-loop subtraction 
is effective at removing the leading lattice artefacts in this quantity. The ratio of
step-scaling functions $\SigmaS/\SigmaP$ provides a second diagnostic check
of our setup. The continuum limit step scaling functions $\sigmaS$ and
$\sigmaP$ are expected to be equal in a $\chi$SF setup. We confirm that
the ratio $\SigmaS/\SigmaP$ fulfils this expectation.

Our main result consists in the computation of the step-scaling function 
$\sigmaP$ (equivalently $\tau/\beta$) from hadronic to electroweak energy scales.
In the high energy regime, we computed the quark mass 
running factor from the switching
scale to the RG-invariant definition of the quark mass.
In the low energy regime, we computed the mass running factor from the hadronic
scale $\mu_{\text{had}}$ to the switching scale. Putting these results together we obtain the total running factor
\begin{equation}
\frac{M}{\mbar(\mu_{\text{had}})} = 0.9108(78) \,\, .
\end{equation}
from eq.(\ref{eq:M-barm-SF}) and
\begin{equation}
\frac{M}{\mbar(\mu_{\text{had}})} = 0.9107(80) \,\, .
\end{equation}
from eq.(\ref{eq:M-barm-SF2}).
These can be compared to the  SF-LQCD result from ref.~\cite{Campos:2018ahf}, namely
$M/\mbar(\mu_{\text{had}}) = 0.9148(88)$, obtained on the
same configuration ensembles.

Our results for the mass running factors are consistent with the findings
of~\cite{Campos:2018ahf}. The two formulations are formally equivalent
in the continuum, but are obtained from two different regularisations
of the valence quark action. Their compatibility is a non-trivial check of 
universality of the two lattice theories. 

Having validated our setup 
quantitatively lays the groundwork for studies of other bilinear operators,
such as the tensor which, without an improvement scheme in SF,
suffers $O(a)$ artefacts. Work towards obtaining automatically improved 
tensor matrix elements is under way; see ref.~\cite{Plasencia:2021ihd}
for preliminary results. Similar advantages are expected for more
complicated four-quark operators, like those used in studies of kaon mixing
beyond the Standard Model. The strategy for computing  automatically 
$O(a)$-improved $B_{\rm\scriptscriptstyle K}$ matrix elements, including BSM contributions, 
in a $\chi$SF renormalisation scheme, has been outlined in refs.~\cite{Mainar:2016uwb,Campos:2019nus}.

\section*{Acknowledgements}
\label{sec:acknowl}
We wish to thank Patrick Fritzsch, Carlos Pena, David Preti, and Alberto Ramos for their help. 
This work  is partially supported by INFN and CINECA, as part of research project of the QCDLAT INFN-initiative. 
We acknowledge the Santander Supercomputacion support group at the University of Cantabria which provided access to the Altamira Supercomputer at the Institute of Physics of Cantabria (IFCA-CSIC).
We also acknowledge support by the Poznan Supercomputing and Networking Center (PSNC) under the project with grant number 466.
AL acknowledges support by the U.S. Department of Energy under grant number DE-SC0015655.

\begin{appendix}
\section{Determination of $\zf$} 
\label{sec:zf}
To obtain ``automatic" $O(a)$-improvement we tune nonperturbatively 
the boundary counterterm $\zf$ in order to satisfy Eq.~\eqref{eq:tune-zf}.
This is done at the $\hopc$ values obtained in
ref.~\cite{Campos:2018ahf} with SF boundary conditions for the quarks fields.
We use an iterative procedure. Starting from an initial guess 
($\zf^{(0)}$, $s^{(0)}$), for $\zf$ and 
$s \equiv \frac{\partial}{\partial \zf} \gA^{ud}(T/2)$, 
we compute $\gA^{ud}(T/2)[\zf] \equiv \gA^{ud}[\zf]$
for a given gauge field ensemble. We then update the values
\begin{align}
\zf^{(i+1)} &= \zf^{(i)} -\gA^{ud}[\zf^{(i)}]/s^{(i)}  \label{eq:zf_update}\\
s^{(i+1)} &= 
(\gA^{ud}[\zf^{(i+1)}] - \gA^{ud}[\zf^{(i)}])/(\zf^{(i+1)}- \zf^{(i)}) 
\label{eq:s_update}\,,
\end{align}
where $\gA^{ud}[\zf^{(i+1)}]$ is computed after~\eqref{eq:zf_update}
in order to update the slope in~\eqref{eq:s_update}.
As convergence criterion we require that the correlator be zero 
within statistical errors, 
\begin{equation} \label{conv_crit}
|\gA^{ud}(T/2) | < \delta \gA^{ud}(T/2) \,.
\end{equation}

\begin{figure}
\includegraphics[width=0.5\textwidth]{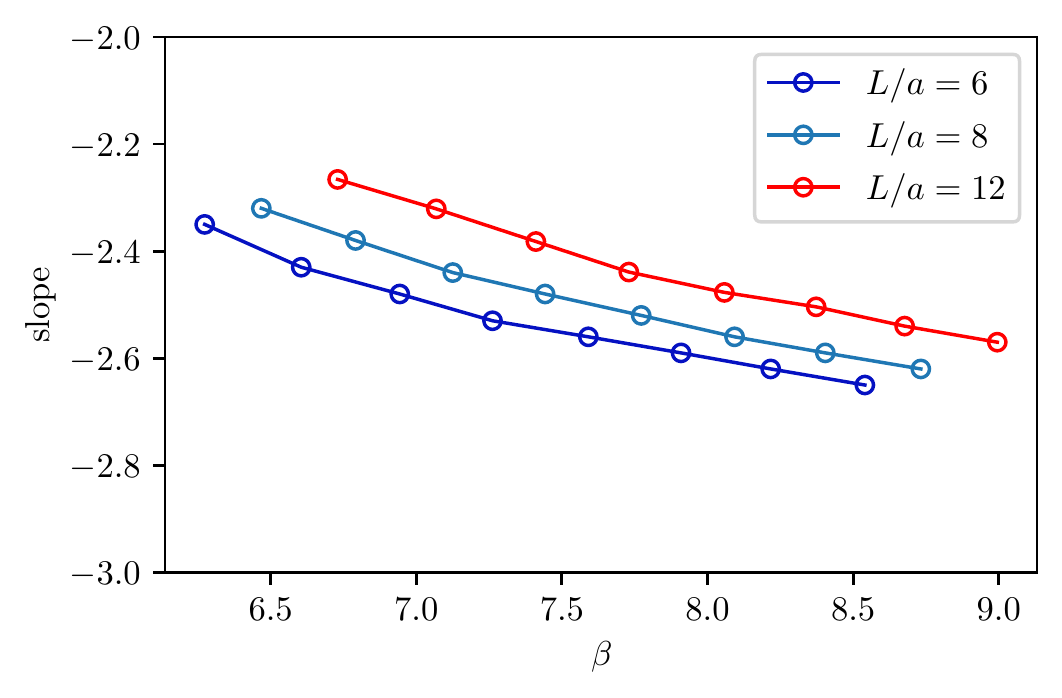}
\caption{The slope $\frac{\partial}{\partial \zf} g_A^{ud}(T/2)$
as a function of $\beta$, with varying $L/a$, in the high energy
range.
\label{fig:slope}}
\end{figure}

As an initial starting guess we take the $O(g_0^2)$ perturbative 
result~\cite{Brida:2016rmy} 
for $\zf^{(0)} = 1+g_0^2 C_F \times 0.16759(1) $.
The initial guess for the slope $s^{(0)}$ was determined empirically
by measuring $\gA^{ud}[\zf]$ for a few $\zf$ values near $\zf^{(0)}$ on
a single ensemble, from which we obtained $s^{(0)} = -2.3$.
In practice the slope is found to be a slowly varying function of $g_0^2$
(cf.\ Fig.~\ref{fig:slope}) and the termination 
of the algorithm does not depend sensitively on its initial value.
Because we are working in a narrow
range around the final value of $\zf$, the correlator $\gA^{ud}[\zf]$ varies
nearly linearly with $\zf$.

For the computation of $\gA^{ud}$
the algorithm is first run in a low-precision mode using 
1000 configurations, and when the convergence criterion~\eqref{conv_crit}
is satisfied, it switches to a high-precision mode using the full ensemble.
The new starting values are the the $\zf$- and the slope-estimates of 
the low-precision run.
The values of $\gA^{ud}[\zf]$ for successive $\zf$ estimates are highly
correlated, so the slopes are determined precisely and the 
algorithm terminates quickly.

Results for both the SF and GF ensembles are shown in
Fig.~\ref{fig:zf_vs_beta}. We see that $\zf$ varies smoothly
with $g_0^2$ in each energy regime (SF and GF). Compared
to the $O(g_0^2)$ perturbative result, it clearly has sizeable
contributions from higher orders.
The results for $\zf$ could turn out to be useful in future studies
of $\chi$SF-LQCD with $\NF=3$. For this reason we present
detailed results on the quantities relevant to this tuning
in a separate Appendix~\ref{app:zfTables}.

\section{Retuning $\hopc$} 
\label{sec:kappa}

The value of $\hopc$ tuned in the $\chi$SF regularisation differs from the value tuned in the SF one by $O(a)$ lattice artefacts. However, when physical observables in the $\chi$SF setup are computed using these two $\hopc$ values, the two determinations  differ only by $O(a^2)$ lattice artefacts~\cite{Brida:2016rmy}. In this Appendix we check that 
physical results computed with $\zf$ tuned with $\hopc$ fixed at the SF values of ref.~\cite{Campos:2018ahf} are compatible 
to those obtained when $\zf$ and $\hopc$ are tuned simultaneously with an iterative procedure in the $\chi$SF setup. 

Our test consists in computing the SSFs of the pseudoscalar and scalar operators \eqref{eq:Ssf-PchiSF} and \eqref{eq:Ssf-SchiSF}, at a single value of the renormalised coupling, $\uSF(\mu_0)$=2.012, for which an extra fine lattice with $L/a = 16$ is available.
Considerations based on $O(a)$-improvement in ref~\cite{DallaBrida:2018tpn} imply that the quark mass depends weakly on $\zf$. This suggests an iterative procedure, in which either $\zf$ or $\hopc$ is tuned in alternation, while the other parameter is held fixed. The output of a tuning stage (say, $\zf$) is thus kept fixed in the successive stage, in which the other parameter (say $\hopc$) is being tuned. As we do not expect the quark mass to change appreciably with small variations of $\zf$, the overall process should converge rapidly.

\begin{figure}[t!]
\includegraphics[width=0.5\textwidth]{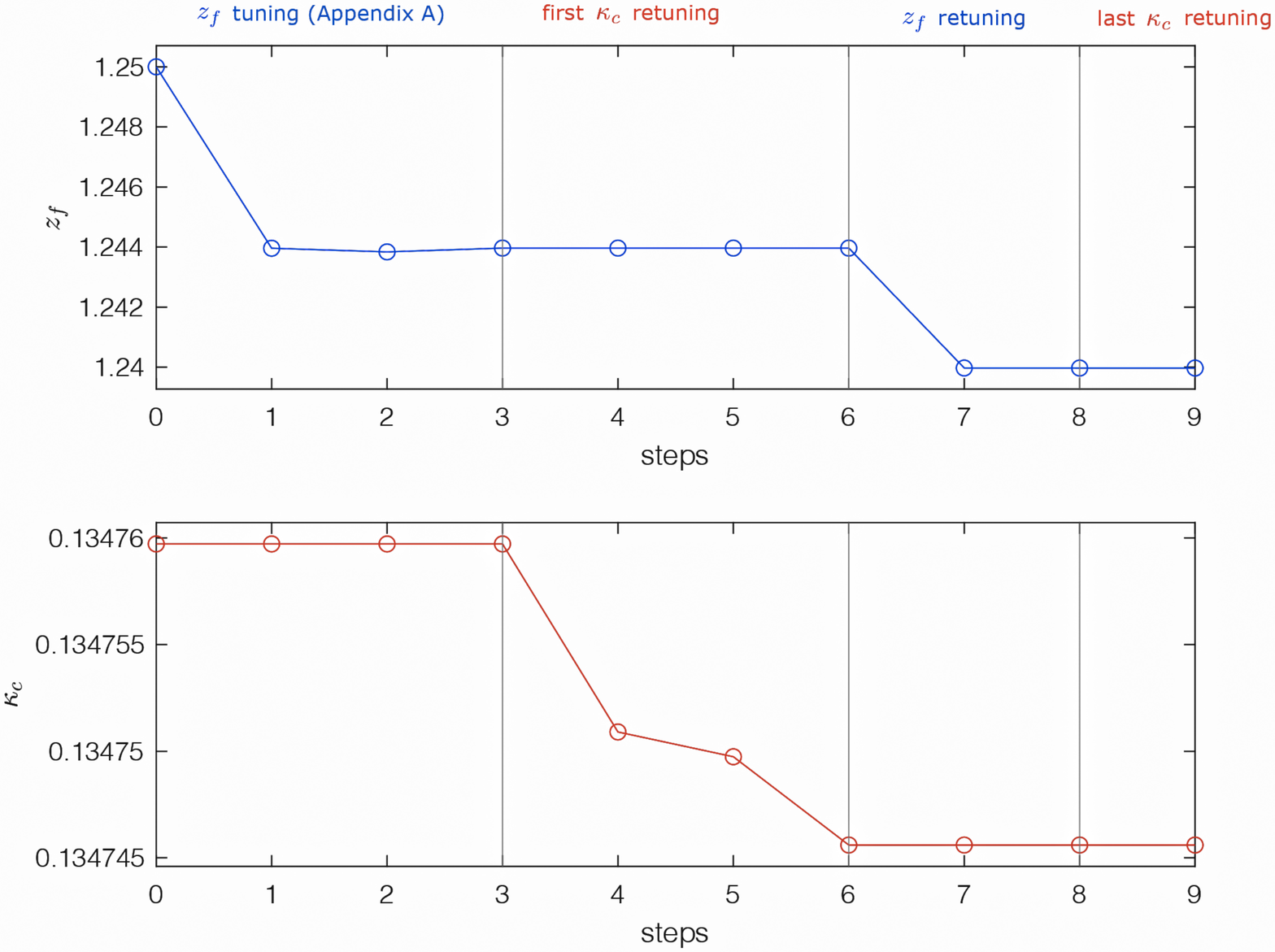}
\caption{Simultaneous tuning of $\zf$ and $\kappa$ at $\uSF$=2.012 and lattice volume $L/a=12$. 
  On the $x$ axis we enumerate the iteration step of the overall tuning procedure; the vertical dotted lines indicate the end of a tuning stage. Each tuning stage ends when the appropriate convergence criterion is met. 
\label{fig:doub_tuning}}
\end{figure}

This procedure is displayed in Fig.~\ref{fig:doub_tuning}. The first stage consists in tuning of $\zf$ with $\hopc$ held fixed at the SF value of ref.~\cite{Campos:2018ahf}. Essentially this is what is described in Appendix~\ref{sec:zf}. The convergence criterion is met after 3 iterations (step 3 in Fig.~\ref{fig:doub_tuning}). Then the second stage begins, where $\hopc$ is tuned with $\zf$ held fixed. This is analogous to what is done in Appendix~\ref{sec:zf}, with $\zf$ replaced by $\hopc$: the initial guesses for $\hopc$ and the slope $\frac{\partial (mL)}{\partial\kappa}$ are taken from refs.~\cite{Campos:2018ahf} and Appendix~B of ref.~\cite{DallaBrida:2018tpn} respectively.
The algorithm first runs in low precision mode and then in high precision mode; after 3 iterations (step 6 in Fig.~\ref{fig:doub_tuning}) the PCAC mass is naught within statistical precision.
Note that in practice we tune the bare quark mass $m_{0}$ using the slope $\frac{\partial (mL)}{\partial (m_{0}L)}$
and then we convert it to $\kappa=\frac{1}{2 m_{0} +8}$. We prefer this because $m$ is linear in $m_{0}$. See ref.~\cite{DallaBrida:2018tpn} for more details.

We can carry on by retuning $\zf$ while keeping  $\hopc$ fixed to its new value; this is stage 3 of our procedure. The new initial guesses for $\zf$ and the slope $\frac{\partial}{\partial \zf} g_A^{ud}(T/2)$ are the output of the previous $\zf$ tuning (stage 1). We  alternate the two tuning procedures until both parameters remain stable within their errors.
\begin{table}
\begin{tabular}{c c | c c}
 $L/a$ & $N_{conf}$ & $ \frac{\zf^{\chi\rm\scriptscriptstyle{SF}} - \zf^{\rm\scriptscriptstyle{SF}} }{\zf^{\chi\rm\scriptscriptstyle{SF}}}\!\times\! 100 $ & $ \frac{\hopc^{\chi\rm\scriptscriptstyle{SF}} - \hopc^{\rm\scriptscriptstyle{SF}}}{\hopc^{\chi\rm\scriptscriptstyle{SF}}}\!\times\! 100 $\\
\hline 
 6 & 5000 & 0.94 & 0.06\\
 8 & 5000 & 0.36 & 0.02 \\
 12 & 3000 & 0.31 & 0.01\\
 16 & 4604 & 0.19 & 0.005\\
\end{tabular}
\caption{Percentage variation of $\zf$ and $\hopc$ (for $\uSF = 2.012$ ensembles)
between the values tuned as in Appendix A (denoted by a superscript SF) and those tuned according to the present procedure (denoted by a superscript $\chi$SF). 
\label{tab:perc_var}}
\end{table}

Looking at the percentage variations of $\zf$ and $\hopc$ after their retuning, we see that both of them have not changed considerably: the variation of $\zf$ is less than 1 percent for all lattice volumes and that of $\hopc$ is smaller than 1 per mil. The values are given in Table \ref{tab:perc_var}.
\begin{figure}[t!]
\includegraphics[width=0.5\textwidth]{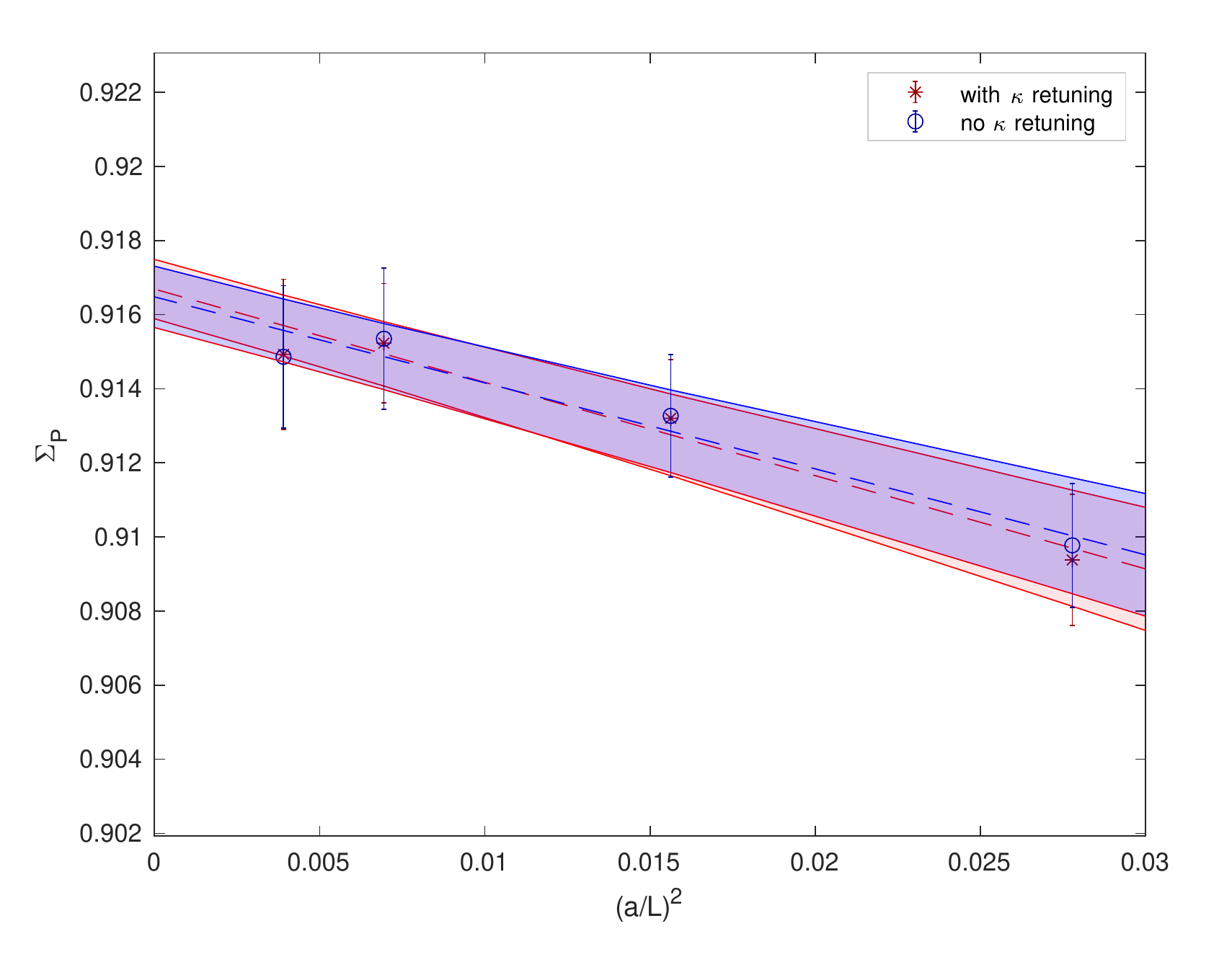}
\caption{$\SigmaPchiSF$ vs $(a/L)^2$ at $\uSF$=2.012. In red we show data and related fits computed with the simultaneous tuning of $\zf$ and $\hopc$; in blue, those computed w/out $\hopc $ retuning.
\label{fig:compare_sigP}}
\end{figure}

We finally compare results for $\SigmaPchiSF$, obtained with and without retuning of $\hopc$. In  
Fig.~\ref{fig:compare_sigP} 
$\SigmaPchiSF$ is plotted against $(a/L)^2$ at the renormalised squared coupling $\uSF$=2.012. The figure shows that  the two sets of data overlap strongly, both at finite lattice spacing and in the continuum.

We conclude that the tuning procedure can be stopped after completion of the first stage (step 3 in Fig.~\ref{fig:doub_tuning}), as described in Appendix~\ref{sec:zf}, without loss of precision for the quantities of interest.

\section{The effect of $d_s$}
\label{sec:ds_check}

\begin{figure}[t!]
\includegraphics[width=0.5\textwidth]{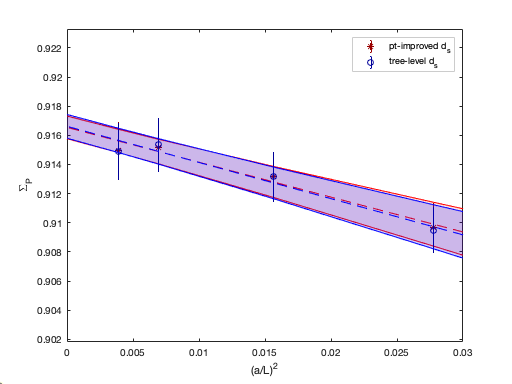}
\caption{
$\SigmaPchiSF$vs $(a/L)^{2}$ at $u_{SF}=2.012$. In red
we show data and related fits computed with $d_{s}^{\rm\scriptscriptstyle tree}$; in
blue, those computed with $d_{s}^{\rm\scriptscriptstyle 1-loop}$.
\label{fig:ds_check}}
\end{figure}

We compare the results for $\SigmaPchiSF$ obtained with the tree level value of $d_s$
to those obtained with the 1-loop result. We perform the test at $u_{SF}$=2.0120, where the difference 
between the tree-level and 1-loop estimates of $d_s$ is the biggest possible in our high energy (SF) range:
($d_{s}^{\rm\scriptscriptstyle 1-loop}-d_{s}^{\rm\scriptscriptstyle tree})=d_{s}^{(1)}\times 2.0120=0.0016(8)$.
As expected and shown in Fig.~\ref{fig:ds_check}, the datapoints
obtained with the two $d_s$ estimates overlap completely.

\section{Simulation details} 
\label{app:zfTables}

\pagebreak 

\providecommand{\tabularnewline}{\\}

\begin{table}

\begin{tabular}{|ccccccccc|}
\hline 
$u$ & $L/a$ & $\beta$ & $\hopc$ & $\icsw$ & $N_{\rm\scriptscriptstyle ms}$ & $\gA^{ud}$ & $\partial \gA^{ud}/\partial \zf$ & $\zf$\tabularnewline
\hline 
\noalign{\vskip\doublerulesep}
\hline 
1.110000 & 6 & 8.5403 & 0.13233610 & 1.233045285565058 & 5000 & 3.34e-07 +/- 0.000529 & -2.6411749 & 1.18588709299\tabularnewline[\doublerulesep]
1.110000 & 8 & 8.7325 & 0.13213380 & 1.224666388699756 & 5000 & -6.87e-07 +/- 0.000397 & -2.6152159 & 1.18374688510\tabularnewline[\doublerulesep]
1.110000 & 12 & 8.9950 & 0.13186210 & 1.214293680665697 & 2769 & 1.89e-06 +/- 0.000348 & -2.5572039 & 1.17624612625\tabularnewline[\doublerulesep]
\hline 
1.184460 & 6 & 8.2170 & 0.13269030 & 1.248924515099129 & 5000 & -1.33e-05 +/- 0.000571 & -2.6258675 & 1.19427949923\tabularnewline[\doublerulesep]
1.184460 & 8 & 8.4044 & 0.13247670 & 1.239426196162344 & 5000 & 7.34e-06 +/- 0.000419 & -2.5982674 & 1.19224223013\tabularnewline[\doublerulesep]
1.184460 & 12 & 8.6769 & 0.13217153 & 1.22701700000000 & 2476 & 1.12e-07 +/- 0.000403 & -2.5383549 & 1.18417905314\tabularnewline[\doublerulesep]
\hline 
1.265690 & 6 & 7.9091 & 0.13305720 & 1.266585617959733 & 5000 & 7.17e-06 +/- 0.000598 & -2.6054349 & 1.20212085045\tabularnewline[\doublerulesep]
1.265690 & 8 & 8.0929 & 0.13283120 & 1.255711356539447 & 5000 & -9.56e-07 +/- 0.000456 & -2.5579952 & 1.20026752592\tabularnewline[\doublerulesep]
1.265690 & 12 & 8.3730 & 0.13249231 & 1.24095900000000 & 2729 & 2.36e-08 +/- 0.000400 & -2.5034648 & 1.19163887560\tabularnewline[\doublerulesep]
\hline 
1.362700 & 6 & 7.5909 & 0.13346930 & 1.288146969458134 & 5000 & -8.57e-07 +/- 0.000656 & -2.5642314 & 1.21157164130\tabularnewline[\doublerulesep]
1.362700 & 8 & 7.7723 & 0.13322830 & 1.275393611340024 & 5000 & -1.20e-06 +/- 0.000485 & -2.5192879 & 1.21009426577\tabularnewline[\doublerulesep]
1.362700 & 12 & 8.0578 & 0.13285365 & 1.25770900000000 & 2448 & 1.20e-06 +/- 0.000456 & -2.4722959 & 1.20100624217\tabularnewline[\doublerulesep]
\hline 
1.480800 & 6 & 7.2618 & 0.13393370 & 1.315030958783770 & 5000 & 1.37e-06 +/- 0.000709 & -2.5311118 & 1.22119155568\tabularnewline[\doublerulesep]
1.480800 & 8 & 7.4424 & 0.13367450 & 1.299622821237046 & 5000 & 2.26e-06 +/- 0.000538 & -2.4850676 & 1.22025587648\tabularnewline[\doublerulesep]
1.480800 & 12 & 7.7299 & 0.13326353 & 1.278252758659668 & 2711 & -5.95e-07 +/- 0.000463 & -2.4422897 & 1.21049262680\tabularnewline[\doublerulesep]
\hline 
1.617300 & 6 & 6.9433 & 0.13442200 & 1.346919223092444 & 5000 & 2.77e-06 +/- 0.000786 & -2.4829699 & 1.23119560164\tabularnewline[\doublerulesep]
1.617300 & 8 & 7.1254 & 0.13414180 & 1.327878356622864 & 5000 & 1.47e-07 +/- 0.000582 & -2.4485296 & 1.23052937460\tabularnewline[\doublerulesep]
1.617300 & 12 & 7.4107 & 0.13369922 & 1.30220600000000 & 2535 & 4.55e-08 +/- 0.000509 & -2.3828505 & 1.22067806449\tabularnewline[\doublerulesep]
\hline 
1.794300 & 6 & 6.6050 & 0.13498290 & 1.389385004928746 & 5000 & -2.33e-06 +/- 0.000859 & -2.4255593 & 1.24169304397\tabularnewline[\doublerulesep]
1.794300 & 8 & 6.7915 & 0.13467650 & 1.364706438701718 & 5000 & -1.19e-05 +/- 0.000638 & -2.3657177 & 1.24218454960\tabularnewline[\doublerulesep]
1.794300 & 12 & 7.0688 & 0.13420891 & 1.333551296494656 & 2339 & -7.03e-06 +/- 0.000592 & -2.3034246 & 1.23170004013\tabularnewline[\doublerulesep]
\hline 
2.012000 & 6 & 6.2735 & 0.13557130 & 1.442967721668930 & 5000 & -8.02e-06 +/- 0.000971 & -2.3627021 & 1.25238467445\tabularnewline[\doublerulesep]
2.012000 & 8 & 6.4680 & 0.13523620 & 1.409845308468962 & 5000 & -3.40e-07 +/- 0.000706 & -2.3251048 & 1.25331910626\tabularnewline[\doublerulesep]
2.012000 & 12 & 6.7299 & 0.13475973 & 1.372481791156670 & 3000 & -1.44e-06 +/- 0.000603 & -2.2547920 & 1.24396571640\tabularnewline[\doublerulesep]
2.012000 & 16 & 6.9346 & 0.13441209 & 1.34788873527000 & 4604 & 3.14e-06 +/- 0.000340 & -2.2150346 & 1.23527136701\tabularnewline[\doublerulesep]
\hline 
\end{tabular}

\caption{\label{tab:SF}
The first column refers to the values of the squared renormalised gauge coupling $\gbar^2(\mu) = u$ in the SF energy region; columns 2 to 5 display  the relevant bare lattice parameters corresponding to $u$; column 6 shows the number of gauge field configurations used for the measurements. The last three columns contain the output of the tuned $\zf$ and the final values of $\gA^{ud}$ and $\partial \gA^{ud}/\partial \zf$.}
\end{table}

\cleardoublepage

\begin{table}
\begin{tabular}{|ccccccccc|}
\hline 
$u$ & $L/a$ & $\beta$ & $\hopc$ & $\icsw$ & $N_{\rm\scriptscriptstyle ms}$ & $\gA^{ud}$ & $\partial \gA^{ud}/\partial \zf$ & $\zf$\tabularnewline
\hline 
\noalign{\vskip\doublerulesep}
\hline 
2.125700 & 8 & 5.3715 & 0.13362120 & 1.259364773796311 & 5000 & 4.06e-06 +/- 0.000534 & -2.4642129 & 1.22654269651\tabularnewline[\doublerulesep]
2.125700 & 12 & 5.5431 & 0.13331407 & 1.244237155112229 & 2001 & -1.15e-06 +/- 0.000285 & -2.3724251 & 1.21985757706\tabularnewline[\doublerulesep]
2.125700 & 16 & 5.7000 & 0.13304840 & 1.232057931661424 & 8000 & 1.70e-06 +/- 0.000217 & -2.2984404 & 1.21389574489\tabularnewline[\doublerulesep]
\hline 
2.390000 & 8 & 5.0710 & 0.13421678 & 1.291712997425573 & 5000 & -2.46e-06 +/- 0.000593 & -2.3937509 & 1.24245430325\tabularnewline[\doublerulesep]
2.390000 & 12 & 5.2425 & 0.13387635 & 1.272228757209511 & 2001 & 1.77e-05 +/- 0.000311 & -2.3527260 & 1.23476121175\tabularnewline[\doublerulesep]
2.390000 & 16 & 5.4000 & 0.13357851 & 1.256705230332892 & 8000 & -1.26e-06 +/- 0.000261 & -2.2530349 & 1.22758469820\tabularnewline[\doublerulesep]
\hline 
2.735900 & 8 & 4.7649 & 0.13488555 & 1.335350323996506 & 5001 & -2.82e-06 +/- 0.000687 & -2.3249628 & 1.25947416858\tabularnewline[\doublerulesep]
2.735900 & 12 & 4.9387 & 0.13450761 & 1.308983384364439 & 2001 & 1.83e-05 +/- 0.000442 & -2.2719935 & 1.25193714648\tabularnewline[\doublerulesep]
2.735900 & 16 & 5.1000 & 0.13416889 & 1.288203306487197 & 5001 & -1.65e-06 +/- 0.000297 & -2.1546738 & 1.24351820011\tabularnewline[\doublerulesep]
\hline 
3.202900 & 8 & 4.4576 & 0.13560675 & 1.395741031275910 & 5001 & 6.60e-05 +/- 0.000793 & -2.3080103 & 1.27801540556\tabularnewline[\doublerulesep]
3.202900 & 12 & 4.6347 & 0.13519986 & 1.358462476494125 & 2001 & 4.79e-06 +/- 0.000520 & -2.1513773 & 1.27063136963\tabularnewline[\doublerulesep]
3.202900 & 16 & 4.8000 & 0.13482139 & 1.329646151978636 & 5001 & 3.06e-07 +/- 0.000360 & -2.0625697 & 1.26149028258\tabularnewline[\doublerulesep]
\hline 
3.864300 & 8 & 4.1519 & 0.13632589 & 1.482418125298923 & 5001 & 3.73e-07 +/- 0.000980 & -2.1399699 & 1.29722920690\tabularnewline[\doublerulesep]
3.864300 & 12 & 4.3317 & 0.13592664 & 1.427424655158656 & 2001 & -9.72e-07 +/- 0.000617 & -2.0302207 & 1.29144247047\tabularnewline[\doublerulesep]
3.864300 & 16 & 4.5000 & 0.13552582 & 1.386110343557152 & 5001 & -1.38e-06 +/- 0.000427 & -1.9484240 & 1.28199124254\tabularnewline[\doublerulesep]
\hline 
4.490100 & 8 & 3.9479 & 0.13674684 & 1.563885414775983 & 5001 & 1.64e-04 +/- 0.001100 & -2.0493154 & 1.30786594013\tabularnewline[\doublerulesep]
4.490100 & 12 & 4.1282 & 0.13640300 & 1.490702297580152 & 2001 & -1.31e-05 +/- 0.000713 & -1.9429389 & 1.30582505718\tabularnewline[\doublerulesep]
4.490100 & 16 & 4.3000 & 0.13600821 & 1.436199798821361 & 5001 & -1.53e-05 +/- 0.000493 & -1.8420700 & 1.29622706156\tabularnewline[\doublerulesep]
\hline 
5.301000 & 8 & 3.7549 & 0.13701929 & 1.668369108400627 & 5001 & -8.19e-06 +/- 0.001380 & -1.9545399 & 1.31461240780\tabularnewline[\doublerulesep]
5.301000 & 12 & 3.9368 & 0.13679805 & 1.569056010619204 & 2001 & 2.31e-05 +/- 0.000831 & -1.8037352 & 1.31770290718\tabularnewline[\doublerulesep]
5.301000 & 16 & 4.1000 & 0.13647301 & 1.500935714848465 & 5001 & -2.01e-05 +/- 0.000717 & -1.7111716 & 1.31005410303\tabularnewline[\doublerulesep]
\hline 
5.867300 & 8 & 3.6538 & 0.13707221 & 1.738234164347418 & 5001 & -3.05e-05 +/- 0.001560 & -1.8809782 & 1.31487229203\tabularnewline[\doublerulesep]
5.867300 & 12 & 3.8333 & 0.13696774 & 1.621966539608638 & 5001 & 5.56e-05 +/- 0.000921 & -1.7457623 & 1.32327345473\tabularnewline[\doublerulesep]
5.867300 & 16 & 4.0000 & 0.13668396 & 1.540714371185832 & 4602 & -2.24e-05 +/- 0.000654 & -1.6560183 & 1.31588724381\tabularnewline[\doublerulesep]
\hline 
6.548900 & 8 & 3.5565 & 0.13703245 & 1.818951161611082 & 5001 & -3.29e-05 +/- 0.001830 & -1.8016743 & 1.31531748561\tabularnewline[\doublerulesep]
6.548900 & 12 & 3.7354 & 0.13708263 & 1.680901952205217 & 5001 & 2.42e-06 +/- 0.001080 & -1.6027921 & 1.32886731276\tabularnewline[\doublerulesep]
6.548900 & 16 & 3.9000 & 0.13687202 & 1.586881030973021 & 4600 & -5.17e-07 +/- 0.000737 & -1.5210420 & 1.32248784097\tabularnewline[\doublerulesep]
\hline 
\end{tabular}

\caption{\label{tab:GF}
The first column refers to the values of the squared renormalised gauge coupling $\gbar^2(\mu) = u$ in the GF energy region; columns 2 to 5 display the relevant bare lattice parameters corresponding to $u$; column 6 shows the number of gauge field configurations used for the measurements. The last three columns contain the output of the tuned $\zf$ and the final values of $\gA^{ud}$ and $\partial \gA^{ud}/\partial \zf$.}

\end{table}

\end{appendix}

\bibliography{paper}

\begin{thebibliography}{48}
\expandafter\ifx\csname natexlab\endcsname\relax\def\natexlab#1{#1}\fi
\expandafter\ifx\csname bibnamefont\endcsname\relax
  \def\bibnamefont#1{#1}\fi
\expandafter\ifx\csname bibfnamefont\endcsname\relax
  \def\bibfnamefont#1{#1}\fi
\expandafter\ifx\csname citenamefont\endcsname\relax
  \def\citenamefont#1{#1}\fi
\expandafter\ifx\csname url\endcsname\relax
  \def\url#1{\texttt{#1}}\fi
\expandafter\ifx\csname urlprefix\endcsname\relax\def\urlprefix{URL }\fi
\providecommand{\bibinfo}[2]{#2}
\providecommand{\eprint}[2][]{\url{#2}}

\bibitem[{\citenamefont{de~Divitiis et~al.}(2019)\citenamefont{de~Divitiis,
  Fritzsch, Heitger, {K\"oster}, Kuberski, and Vladikas}}]{deDivitiis:2019xla}
\bibinfo{author}{\bibfnamefont{G.~M.} \bibnamefont{de~Divitiis}},
  \bibinfo{author}{\bibfnamefont{P.}~\bibnamefont{Fritzsch}},
  \bibinfo{author}{\bibfnamefont{J.}~\bibnamefont{Heitger}},
  \bibinfo{author}{\bibfnamefont{C.~C.} \bibnamefont{{K\"oster}}},
  \bibinfo{author}{\bibfnamefont{S.}~\bibnamefont{Kuberski}}, \bibnamefont{and}
  \bibinfo{author}{\bibfnamefont{A.}~\bibnamefont{Vladikas}}
  (\bibinfo{collaboration}{ALPHA}), \bibinfo{journal}{Eur. Phys. J.}
  \textbf{\bibinfo{volume}{C79}}, \bibinfo{pages}{797} (\bibinfo{year}{2019}),
  \eprint{1906.03445}.

\bibitem[{\citenamefont{Heitger et~al.}(2020)\citenamefont{Heitger, Joswig, and
  Vladikas}}]{Heitger:2020mkp}
\bibinfo{author}{\bibfnamefont{J.}~\bibnamefont{Heitger}},
  \bibinfo{author}{\bibfnamefont{F.}~\bibnamefont{Joswig}}, \bibnamefont{and}
  \bibinfo{author}{\bibfnamefont{A.}~\bibnamefont{Vladikas}}
  (\bibinfo{collaboration}{ALPHA}), \bibinfo{journal}{Eur. Phys. J. C}
  \textbf{\bibinfo{volume}{80}}, \bibinfo{pages}{765} (\bibinfo{year}{2020}),
  \eprint{2005.01352}.

\bibitem[{\citenamefont{Heitger et~al.}(2021)\citenamefont{Heitger, Joswig,
  Petrak, and Vladikas}}]{Heitger:2021bmg}
\bibinfo{author}{\bibfnamefont{J.}~\bibnamefont{Heitger}},
  \bibinfo{author}{\bibfnamefont{F.}~\bibnamefont{Joswig}},
  \bibinfo{author}{\bibfnamefont{P.~L.~J.} \bibnamefont{Petrak}},
  \bibnamefont{and} \bibinfo{author}{\bibfnamefont{A.}~\bibnamefont{Vladikas}},
  \bibinfo{journal}{Eur. Phys. J. C} \textbf{\bibinfo{volume}{81}},
  \bibinfo{pages}{606} (\bibinfo{year}{2021}), \eprint{2101.10969}.

\bibitem[{\citenamefont{Campos et~al.}(2018)\citenamefont{Campos, Fritzsch,
  Pena, Preti, Ramos, and Vladikas}}]{Campos:2018ahf}
\bibinfo{author}{\bibfnamefont{I.}~\bibnamefont{Campos}},
  \bibinfo{author}{\bibfnamefont{P.}~\bibnamefont{Fritzsch}},
  \bibinfo{author}{\bibfnamefont{C.}~\bibnamefont{Pena}},
  \bibinfo{author}{\bibfnamefont{D.}~\bibnamefont{Preti}},
  \bibinfo{author}{\bibfnamefont{A.}~\bibnamefont{Ramos}}, \bibnamefont{and}
  \bibinfo{author}{\bibfnamefont{A.}~\bibnamefont{Vladikas}},
  \bibinfo{journal}{Eur. Phys. J.} \textbf{\bibinfo{volume}{C78}},
  \bibinfo{pages}{387} (\bibinfo{year}{2018}), \eprint{1802.05243}.

\bibitem[{\citenamefont{Sint}(2011)}]{Sint:2010eh}
\bibinfo{author}{\bibfnamefont{S.}~\bibnamefont{Sint}}, \bibinfo{journal}{Nucl.
  Phys.} \textbf{\bibinfo{volume}{B847}}, \bibinfo{pages}{491}
  (\bibinfo{year}{2011}), \eprint{1008.4857}.

\bibitem[{\citenamefont{{L\"uscher} et~al.}(1996)\citenamefont{{L\"uscher},
  Sint, Sommer, and Weisz}}]{Luscher:1996sc}
\bibinfo{author}{\bibfnamefont{M.}~\bibnamefont{{L\"uscher}}},
  \bibinfo{author}{\bibfnamefont{S.}~\bibnamefont{Sint}},
  \bibinfo{author}{\bibfnamefont{R.}~\bibnamefont{Sommer}}, \bibnamefont{and}
  \bibinfo{author}{\bibfnamefont{P.}~\bibnamefont{Weisz}},
  \bibinfo{journal}{Nucl. Phys.} \textbf{\bibinfo{volume}{B478}},
  \bibinfo{pages}{365} (\bibinfo{year}{1996}), \eprint{hep-lat/9605038}.

\bibitem[{\citenamefont{Dalla~Brida
  et~al.}(2016{\natexlab{a}})\citenamefont{Dalla~Brida, Sint, and
  Vilaseca}}]{Brida:2016rmy}
\bibinfo{author}{\bibfnamefont{M.}~\bibnamefont{Dalla~Brida}},
  \bibinfo{author}{\bibfnamefont{S.}~\bibnamefont{Sint}}, \bibnamefont{and}
  \bibinfo{author}{\bibfnamefont{P.}~\bibnamefont{Vilaseca}},
  \bibinfo{journal}{JHEP} \textbf{\bibinfo{volume}{08}}, \bibinfo{pages}{102}
  (\bibinfo{year}{2016}{\natexlab{a}}), \eprint{1603.00046}.

\bibitem[{\citenamefont{Sint and Leder}(2010)}]{Sint:2010xy}
\bibinfo{author}{\bibfnamefont{S.}~\bibnamefont{Sint}} \bibnamefont{and}
  \bibinfo{author}{\bibfnamefont{B.}~\bibnamefont{Leder}},
  \bibinfo{journal}{PoS} \textbf{\bibinfo{volume}{LATTICE2010}},
  \bibinfo{pages}{265} (\bibinfo{year}{2010}), \eprint{1012.2500}.

\bibitem[{\citenamefont{Sommer}(1994)}]{Sommer:1993ce}
\bibinfo{author}{\bibfnamefont{R.}~\bibnamefont{Sommer}},
  \bibinfo{journal}{Nucl. Phys. B} \textbf{\bibinfo{volume}{411}},
  \bibinfo{pages}{839} (\bibinfo{year}{1994}), \eprint{hep-lat/9310022}.

\bibitem[{\citenamefont{Lopez et~al.}(2013{\natexlab{a}})\citenamefont{Lopez,
  Jansen, Renner, and Shindler}}]{Lopez:2012as}
\bibinfo{author}{\bibfnamefont{J.}~\bibnamefont{Lopez}},
  \bibinfo{author}{\bibfnamefont{K.}~\bibnamefont{Jansen}},
  \bibinfo{author}{\bibfnamefont{D.}~\bibnamefont{Renner}}, \bibnamefont{and}
  \bibinfo{author}{\bibfnamefont{A.}~\bibnamefont{Shindler}},
  \bibinfo{journal}{Nucl. Phys. B} \textbf{\bibinfo{volume}{867}},
  \bibinfo{pages}{567} (\bibinfo{year}{2013}{\natexlab{a}}),
  \eprint{1208.4591}.

\bibitem[{\citenamefont{Lopez et~al.}(2013{\natexlab{b}})\citenamefont{Lopez,
  Jansen, Renner, and Shindler}}]{Lopez:2012mc}
\bibinfo{author}{\bibfnamefont{J.}~\bibnamefont{Lopez}},
  \bibinfo{author}{\bibfnamefont{K.}~\bibnamefont{Jansen}},
  \bibinfo{author}{\bibfnamefont{D.}~\bibnamefont{Renner}}, \bibnamefont{and}
  \bibinfo{author}{\bibfnamefont{A.}~\bibnamefont{Shindler}},
  \bibinfo{journal}{Nucl. Phys. B} \textbf{\bibinfo{volume}{867}},
  \bibinfo{pages}{609} (\bibinfo{year}{2013}{\natexlab{b}}),
  \eprint{1208.4661}.

\bibitem[{\citenamefont{Capitani et~al.}(1999)\citenamefont{Capitani, L{\"
  u}scher, Sommer, and Wittig}}]{Capitani:1998mq}
\bibinfo{author}{\bibfnamefont{S.}~\bibnamefont{Capitani}},
  \bibinfo{author}{\bibfnamefont{M.}~\bibnamefont{L{\" u}scher}},
  \bibinfo{author}{\bibfnamefont{R.}~\bibnamefont{Sommer}}, \bibnamefont{and}
  \bibinfo{author}{\bibfnamefont{H.}~\bibnamefont{Wittig}},
  \bibinfo{journal}{Nucl.\ Phys.\ B} \textbf{\bibinfo{volume}{544}},
  \bibinfo{pages}{669} (\bibinfo{year}{1999}), \bibinfo{note}{[Erratum:
  Nucl.Phys.B 582, 762--762 (2000)]}, \eprint{hep-lat/9810063}.

\bibitem[{\citenamefont{Dalla~Brida et~al.}(2019)\citenamefont{Dalla~Brida,
  Korzec, Sint, and Vilaseca}}]{DallaBrida:2018tpn}
\bibinfo{author}{\bibfnamefont{M.}~\bibnamefont{Dalla~Brida}},
  \bibinfo{author}{\bibfnamefont{T.}~\bibnamefont{Korzec}},
  \bibinfo{author}{\bibfnamefont{S.}~\bibnamefont{Sint}}, \bibnamefont{and}
  \bibinfo{author}{\bibfnamefont{P.}~\bibnamefont{Vilaseca}},
  \bibinfo{journal}{Eur. Phys. J.} \textbf{\bibinfo{volume}{C79}},
  \bibinfo{pages}{23} (\bibinfo{year}{2019}), \eprint{1808.09236}.

\bibitem[{\citenamefont{Dimopoulos et~al.}(2008)\citenamefont{Dimopoulos,
  Herdoiza, Palombi, Papinutto, Pena, Vladikas, and
  Wittig}}]{Dimopoulos:2007ht}
\bibinfo{author}{\bibfnamefont{P.}~\bibnamefont{Dimopoulos}},
  \bibinfo{author}{\bibfnamefont{G.}~\bibnamefont{Herdoiza}},
  \bibinfo{author}{\bibfnamefont{F.}~\bibnamefont{Palombi}},
  \bibinfo{author}{\bibfnamefont{M.}~\bibnamefont{Papinutto}},
  \bibinfo{author}{\bibfnamefont{C.}~\bibnamefont{Pena}},
  \bibinfo{author}{\bibfnamefont{A.}~\bibnamefont{Vladikas}}, \bibnamefont{and}
  \bibinfo{author}{\bibfnamefont{H.}~\bibnamefont{Wittig}}
  (\bibinfo{collaboration}{ALPHA}), \bibinfo{journal}{JHEP}
  \textbf{\bibinfo{volume}{05}}, \bibinfo{pages}{065} (\bibinfo{year}{2008}),
  \eprint{0712.2429}.

\bibitem[{\citenamefont{Dimopoulos et~al.}(2018)\citenamefont{Dimopoulos,
  Herdo\'\i{}za, Papinutto, Pena, Preti, and Vladikas}}]{Dimopoulos:2018zef}
\bibinfo{author}{\bibfnamefont{P.}~\bibnamefont{Dimopoulos}},
  \bibinfo{author}{\bibfnamefont{G.}~\bibnamefont{Herdo\'\i{}za}},
  \bibinfo{author}{\bibfnamefont{M.}~\bibnamefont{Papinutto}},
  \bibinfo{author}{\bibfnamefont{C.}~\bibnamefont{Pena}},
  \bibinfo{author}{\bibfnamefont{D.}~\bibnamefont{Preti}}, \bibnamefont{and}
  \bibinfo{author}{\bibfnamefont{A.}~\bibnamefont{Vladikas}}
  (\bibinfo{collaboration}{ALPHA}), \bibinfo{journal}{Eur. Phys. J. C}
  \textbf{\bibinfo{volume}{78}}, \bibinfo{pages}{579} (\bibinfo{year}{2018}),
  \eprint{1801.09455}.

\bibitem[{\citenamefont{Dimopoulos et~al.}(2009)\citenamefont{Dimopoulos,
  Simma, and Vladikas}}]{Dimopoulos:2009es}
\bibinfo{author}{\bibfnamefont{P.}~\bibnamefont{Dimopoulos}},
  \bibinfo{author}{\bibfnamefont{H.}~\bibnamefont{Simma}}, \bibnamefont{and}
  \bibinfo{author}{\bibfnamefont{A.}~\bibnamefont{Vladikas}},
  \bibinfo{journal}{JHEP} \textbf{\bibinfo{volume}{07}}, \bibinfo{pages}{007}
  (\bibinfo{year}{2009}), \eprint{0902.1074}.

\bibitem[{\citenamefont{Mainar et~al.}(2016)\citenamefont{Mainar, Dalla~Brida,
  and Papinutto}}]{Mainar:2016uwb}
\bibinfo{author}{\bibfnamefont{P.~V.} \bibnamefont{Mainar}},
  \bibinfo{author}{\bibfnamefont{M.}~\bibnamefont{Dalla~Brida}},
  \bibnamefont{and}
  \bibinfo{author}{\bibfnamefont{M.}~\bibnamefont{Papinutto}},
  \bibinfo{journal}{PoS} \textbf{\bibinfo{volume}{LATTICE2015}},
  \bibinfo{pages}{252} (\bibinfo{year}{2016}).

\bibitem[{\citenamefont{Campos et~al.}(2019)\citenamefont{Campos, Dalla~Brida,
  de~Divitiis, Lytle, Papinutto, and Vladikas}}]{Campos:2019nus}
\bibinfo{author}{\bibfnamefont{I.}~\bibnamefont{Campos}},
  \bibinfo{author}{\bibfnamefont{M.}~\bibnamefont{Dalla~Brida}},
  \bibinfo{author}{\bibfnamefont{G.~M.} \bibnamefont{de~Divitiis}},
  \bibinfo{author}{\bibfnamefont{A.}~\bibnamefont{Lytle}},
  \bibinfo{author}{\bibfnamefont{M.}~\bibnamefont{Papinutto}},
  \bibnamefont{and} \bibinfo{author}{\bibfnamefont{A.}~\bibnamefont{Vladikas}},
  in \emph{\bibinfo{booktitle}{{37th International Symposium on Lattice Field
  Theory}}} (\bibinfo{year}{2019}), \eprint{1910.01898}.

\bibitem[{\citenamefont{Plasencia
  et~al.}(2021{\natexlab{a}})\citenamefont{Plasencia, Brida, de~Divitiis,
  Lytle, Papinutto, Pirelli, and Vladikas}}]{Plasencia:2021gjp}
\bibinfo{author}{\bibfnamefont{I.~C.} \bibnamefont{Plasencia}},
  \bibinfo{author}{\bibfnamefont{M.~D.} \bibnamefont{Brida}},
  \bibinfo{author}{\bibfnamefont{G.~M.} \bibnamefont{de~Divitiis}},
  \bibinfo{author}{\bibfnamefont{A.}~\bibnamefont{Lytle}},
  \bibinfo{author}{\bibfnamefont{M.}~\bibnamefont{Papinutto}},
  \bibinfo{author}{\bibfnamefont{L.}~\bibnamefont{Pirelli}}, \bibnamefont{and}
  \bibinfo{author}{\bibfnamefont{A.}~\bibnamefont{Vladikas}}, in
  \emph{\bibinfo{booktitle}{{38th International Symposium on Lattice Field
  Theory}}} (\bibinfo{year}{2021}{\natexlab{a}}), \eprint{2111.15384}.

\bibitem[{\citenamefont{Frezzotti and Rossi}(2004)}]{Frezzotti:2003ni}
\bibinfo{author}{\bibfnamefont{R.}~\bibnamefont{Frezzotti}} \bibnamefont{and}
  \bibinfo{author}{\bibfnamefont{G.}~\bibnamefont{Rossi}},
  \bibinfo{journal}{JHEP} \textbf{\bibinfo{volume}{08}}, \bibinfo{pages}{007}
  (\bibinfo{year}{2004}), \eprint{hep-lat/0306014}.

\bibitem[{\citenamefont{Sint}(2007)}]{Sint:2007ug}
\bibinfo{author}{\bibfnamefont{S.}~\bibnamefont{Sint}}, in
  \emph{\bibinfo{booktitle}{{Workshop on Perspectives in Lattice QCD}}}
  (\bibinfo{year}{2007}), \eprint{hep-lat/0702008}.

\bibitem[{\citenamefont{Dalla~Brida
  et~al.}(2016{\natexlab{b}})\citenamefont{Dalla~Brida, Fritzsch, Korzec,
  Ramos, Sint, and Sommer}}]{Brida:2016flw}
\bibinfo{author}{\bibfnamefont{M.}~\bibnamefont{Dalla~Brida}},
  \bibinfo{author}{\bibfnamefont{P.}~\bibnamefont{Fritzsch}},
  \bibinfo{author}{\bibfnamefont{T.}~\bibnamefont{Korzec}},
  \bibinfo{author}{\bibfnamefont{A.}~\bibnamefont{Ramos}},
  \bibinfo{author}{\bibfnamefont{S.}~\bibnamefont{Sint}}, \bibnamefont{and}
  \bibinfo{author}{\bibfnamefont{R.}~\bibnamefont{Sommer}}
  (\bibinfo{collaboration}{ALPHA}), \bibinfo{journal}{Phys. Rev. Lett.}
  \textbf{\bibinfo{volume}{117}}, \bibinfo{pages}{182001}
  (\bibinfo{year}{2016}{\natexlab{b}}), \eprint{1604.06193}.

\bibitem[{\citenamefont{Dalla~Brida et~al.}(2017)\citenamefont{Dalla~Brida,
  Fritzsch, Korzec, Ramos, Sint, and Sommer}}]{DallaBrida:2016kgh}
\bibinfo{author}{\bibfnamefont{M.}~\bibnamefont{Dalla~Brida}},
  \bibinfo{author}{\bibfnamefont{P.}~\bibnamefont{Fritzsch}},
  \bibinfo{author}{\bibfnamefont{T.}~\bibnamefont{Korzec}},
  \bibinfo{author}{\bibfnamefont{A.}~\bibnamefont{Ramos}},
  \bibinfo{author}{\bibfnamefont{S.}~\bibnamefont{Sint}}, \bibnamefont{and}
  \bibinfo{author}{\bibfnamefont{R.}~\bibnamefont{Sommer}}
  (\bibinfo{collaboration}{ALPHA}), \bibinfo{journal}{Phys. Rev. D}
  \textbf{\bibinfo{volume}{95}}, \bibinfo{pages}{014507}
  (\bibinfo{year}{2017}), \eprint{1607.06423}.

\bibitem[{\citenamefont{Dalla~Brida et~al.}(2018)\citenamefont{Dalla~Brida,
  Fritzsch, Korzec, Ramos, Sint, and Sommer}}]{DallaBrida:2018rfy}
\bibinfo{author}{\bibfnamefont{M.}~\bibnamefont{Dalla~Brida}},
  \bibinfo{author}{\bibfnamefont{P.}~\bibnamefont{Fritzsch}},
  \bibinfo{author}{\bibfnamefont{T.}~\bibnamefont{Korzec}},
  \bibinfo{author}{\bibfnamefont{A.}~\bibnamefont{Ramos}},
  \bibinfo{author}{\bibfnamefont{S.}~\bibnamefont{Sint}}, \bibnamefont{and}
  \bibinfo{author}{\bibfnamefont{R.}~\bibnamefont{Sommer}}
  (\bibinfo{collaboration}{ALPHA}), \bibinfo{journal}{Eur. Phys. J. C}
  \textbf{\bibinfo{volume}{78}}, \bibinfo{pages}{372} (\bibinfo{year}{2018}),
  \eprint{1803.10230}.

\bibitem[{\citenamefont{{L\"uscher} et~al.}(1992)\citenamefont{{L\"uscher},
  Narayanan, Weisz, and Wolff}}]{Luscher:1992an}
\bibinfo{author}{\bibfnamefont{M.}~\bibnamefont{{L\"uscher}}},
  \bibinfo{author}{\bibfnamefont{R.}~\bibnamefont{Narayanan}},
  \bibinfo{author}{\bibfnamefont{P.}~\bibnamefont{Weisz}}, \bibnamefont{and}
  \bibinfo{author}{\bibfnamefont{U.}~\bibnamefont{Wolff}},
  \bibinfo{journal}{Nucl. Phys. B} \textbf{\bibinfo{volume}{384}},
  \bibinfo{pages}{168} (\bibinfo{year}{1992}), \eprint{hep-lat/9207009}.

\bibitem[{\citenamefont{Luscher et~al.}(1994)\citenamefont{Luscher, Sommer,
  Weisz, and Wolff}}]{Luscher:1993gh}
\bibinfo{author}{\bibfnamefont{M.}~\bibnamefont{Luscher}},
  \bibinfo{author}{\bibfnamefont{R.}~\bibnamefont{Sommer}},
  \bibinfo{author}{\bibfnamefont{P.}~\bibnamefont{Weisz}}, \bibnamefont{and}
  \bibinfo{author}{\bibfnamefont{U.}~\bibnamefont{Wolff}},
  \bibinfo{journal}{Nucl. Phys. B} \textbf{\bibinfo{volume}{413}},
  \bibinfo{pages}{481} (\bibinfo{year}{1994}), \eprint{hep-lat/9309005}.

\bibitem[{\citenamefont{Fritzsch and Ramos}(2013)}]{Fritzsch:2013je}
\bibinfo{author}{\bibfnamefont{P.}~\bibnamefont{Fritzsch}} \bibnamefont{and}
  \bibinfo{author}{\bibfnamefont{A.}~\bibnamefont{Ramos}},
  \bibinfo{journal}{JHEP} \textbf{\bibinfo{volume}{10}}, \bibinfo{pages}{008}
  (\bibinfo{year}{2013}), \eprint{1301.4388}.

\bibitem[{\citenamefont{Bruno et~al.}(2017)\citenamefont{Bruno, Dalla~Brida,
  Fritzsch, Korzec, Ramos, Schaefer, Simma, Sint, and Sommer}}]{Bruno:2017gxd}
\bibinfo{author}{\bibfnamefont{M.}~\bibnamefont{Bruno}},
  \bibinfo{author}{\bibfnamefont{M.}~\bibnamefont{Dalla~Brida}},
  \bibinfo{author}{\bibfnamefont{P.}~\bibnamefont{Fritzsch}},
  \bibinfo{author}{\bibfnamefont{T.}~\bibnamefont{Korzec}},
  \bibinfo{author}{\bibfnamefont{A.}~\bibnamefont{Ramos}},
  \bibinfo{author}{\bibfnamefont{S.}~\bibnamefont{Schaefer}},
  \bibinfo{author}{\bibfnamefont{H.}~\bibnamefont{Simma}},
  \bibinfo{author}{\bibfnamefont{S.}~\bibnamefont{Sint}}, \bibnamefont{and}
  \bibinfo{author}{\bibfnamefont{R.}~\bibnamefont{Sommer}}
  (\bibinfo{collaboration}{ALPHA}), \bibinfo{journal}{Phys. Rev. Lett.}
  \textbf{\bibinfo{volume}{119}}, \bibinfo{pages}{102001}
  (\bibinfo{year}{2017}), \eprint{1706.03821}.

\bibitem[{\citenamefont{Wilson}(1974)}]{Wilson:1974sk}
\bibinfo{author}{\bibfnamefont{K.~G.} \bibnamefont{Wilson}},
  \bibinfo{journal}{Phys. Rev. D} \textbf{\bibinfo{volume}{10}},
  \bibinfo{pages}{2445} (\bibinfo{year}{1974}).

\bibitem[{\citenamefont{Sheikholeslami and
  Wohlert}(1985)}]{Sheikholeslami:1985ij}
\bibinfo{author}{\bibfnamefont{B.}~\bibnamefont{Sheikholeslami}}
  \bibnamefont{and} \bibinfo{author}{\bibfnamefont{R.}~\bibnamefont{Wohlert}},
  \bibinfo{journal}{Nucl. Phys. B} \textbf{\bibinfo{volume}{259}},
  \bibinfo{pages}{572} (\bibinfo{year}{1985}).

\bibitem[{\citenamefont{Yamada et~al.}(2005)}]{Yamada:2004ja}
\bibinfo{author}{\bibfnamefont{N.}~\bibnamefont{Yamada}} \bibnamefont{et~al.}
  (\bibinfo{collaboration}{JLQCD, CP-PACS}), \bibinfo{journal}{Phys. Rev. D}
  \textbf{\bibinfo{volume}{71}}, \bibinfo{pages}{054505}
  (\bibinfo{year}{2005}), \eprint{hep-lat/0406028}.

\bibitem[{\citenamefont{Sint and Weisz}(1997)}]{Sint:1997jx}
\bibinfo{author}{\bibfnamefont{S.}~\bibnamefont{Sint}} \bibnamefont{and}
  \bibinfo{author}{\bibfnamefont{P.}~\bibnamefont{Weisz}},
  \bibinfo{journal}{Nucl. Phys. B} \textbf{\bibinfo{volume}{502}},
  \bibinfo{pages}{251} (\bibinfo{year}{1997}), \eprint{hep-lat/9704001}.

\bibitem[{\citenamefont{Bode et~al.}(1999)\citenamefont{Bode, Wolff, and
  Weisz}}]{Bode:1998hd}
\bibinfo{author}{\bibfnamefont{A.}~\bibnamefont{Bode}},
  \bibinfo{author}{\bibfnamefont{U.}~\bibnamefont{Wolff}}, \bibnamefont{and}
  \bibinfo{author}{\bibfnamefont{P.}~\bibnamefont{Weisz}}
  (\bibinfo{collaboration}{Alpha}), \bibinfo{journal}{Nucl. Phys. B}
  \textbf{\bibinfo{volume}{540}}, \bibinfo{pages}{491} (\bibinfo{year}{1999}),
  \eprint{hep-lat/9809175}.

\bibitem[{\citenamefont{{L\"uscher} and Weisz}(1985)}]{Luscher:1985zq}
\bibinfo{author}{\bibfnamefont{M.}~\bibnamefont{{L\"uscher}}} \bibnamefont{and}
  \bibinfo{author}{\bibfnamefont{P.}~\bibnamefont{Weisz}},
  \bibinfo{journal}{Phys. Lett. B} \textbf{\bibinfo{volume}{158}},
  \bibinfo{pages}{250} (\bibinfo{year}{1985}).

\bibitem[{\citenamefont{Bulava and Schaefer}(2013)}]{Bulava:2013cta}
\bibinfo{author}{\bibfnamefont{J.}~\bibnamefont{Bulava}} \bibnamefont{and}
  \bibinfo{author}{\bibfnamefont{S.}~\bibnamefont{Schaefer}},
  \bibinfo{journal}{Nucl. Phys. B} \textbf{\bibinfo{volume}{874}},
  \bibinfo{pages}{188} (\bibinfo{year}{2013}), \eprint{1304.7093}.

\bibitem[{\citenamefont{Vilaseca}()}]{VilasecaPrivate}
\bibinfo{author}{\bibfnamefont{P.}~\bibnamefont{Vilaseca}},
  \bibinfo{note}{{private communication}}.

\bibitem[{\citenamefont{Aoki et~al.}(1999)\citenamefont{Aoki, Frezzotti, and
  Weisz}}]{Aoki:1998qd}
\bibinfo{author}{\bibfnamefont{S.}~\bibnamefont{Aoki}},
  \bibinfo{author}{\bibfnamefont{R.}~\bibnamefont{Frezzotti}},
  \bibnamefont{and} \bibinfo{author}{\bibfnamefont{P.}~\bibnamefont{Weisz}},
  \bibinfo{journal}{Nucl. Phys. B} \textbf{\bibinfo{volume}{540}},
  \bibinfo{pages}{501} (\bibinfo{year}{1999}), \eprint{hep-lat/9808007}.

\bibitem[{\citenamefont{{G.\ P.\ Lepage}}({\natexlab{a}})}]{lsqfit}
\bibinfo{author}{\bibnamefont{{G.\ P.\ Lepage}}}, \emph{\bibinfo{title}{lsqfit
  (version 11.7)}}, \urlprefix\url{https://github.com/gplepage/lsqfit}.

\bibitem[{\citenamefont{{G.\ P.\ Lepage}}({\natexlab{b}})}]{gvar}
\bibinfo{author}{\bibnamefont{{G.\ P.\ Lepage}}}, \emph{\bibinfo{title}{gvar
  (version 11.9.1)}}, \urlprefix\url{https://github.com/gplepage/gvar}.

\bibitem[{\citenamefont{Wolff}(2004)}]{Wolff:2003sm}
\bibinfo{author}{\bibfnamefont{U.}~\bibnamefont{Wolff}},
  \bibinfo{journal}{Comput. Phys. Commun.} \textbf{\bibinfo{volume}{156}},
  \bibinfo{pages}{143} (\bibinfo{year}{2004}), \bibinfo{note}{[Erratum: Comput.
  Phys. Commun.176,383(2007)]}, \eprint{hep-lat/0306017}.

\bibitem[{\citenamefont{Bruno et~al.}(2015)}]{Bruno:2014jqa}
\bibinfo{author}{\bibfnamefont{M.}~\bibnamefont{Bruno}} \bibnamefont{et~al.},
  \bibinfo{journal}{JHEP} \textbf{\bibinfo{volume}{02}}, \bibinfo{pages}{043}
  (\bibinfo{year}{2015}), \eprint{1411.3982}.

\bibitem[{\citenamefont{Mohler et~al.}(2018)\citenamefont{Mohler, Schaefer, and
  Simeth}}]{Mohler:2017wnb}
\bibinfo{author}{\bibfnamefont{D.}~\bibnamefont{Mohler}},
  \bibinfo{author}{\bibfnamefont{S.}~\bibnamefont{Schaefer}}, \bibnamefont{and}
  \bibinfo{author}{\bibfnamefont{J.}~\bibnamefont{Simeth}},
  \bibinfo{journal}{EPJ Web Conf.} \textbf{\bibinfo{volume}{175}},
  \bibinfo{pages}{02010} (\bibinfo{year}{2018}), \eprint{1712.04884}.

\bibitem[{\citenamefont{Dalla~Brida et~al.}()\citenamefont{Dalla~Brida,
  Mellini, Papinutto, Scardino, and Vilaseca}}]{chiSF-PT-TEMP}
\bibinfo{author}{\bibfnamefont{M.}~\bibnamefont{Dalla~Brida}},
  \bibinfo{author}{\bibfnamefont{A.}~\bibnamefont{Mellini}},
  \bibinfo{author}{\bibfnamefont{M.}~\bibnamefont{Papinutto}},
  \bibinfo{author}{\bibfnamefont{F.}~\bibnamefont{Scardino}}, \bibnamefont{and}
  \bibinfo{author}{\bibfnamefont{P.}~\bibnamefont{Vilaseca}}
  (\bibinfo{collaboration}{ALPHA}), \bibinfo{note}{{in preparation}}.

\bibitem[{\citenamefont{Sint and Weisz}(1999)}]{Sint:1998iq}
\bibinfo{author}{\bibfnamefont{S.}~\bibnamefont{Sint}} \bibnamefont{and}
  \bibinfo{author}{\bibfnamefont{P.}~\bibnamefont{Weisz}}
  (\bibinfo{collaboration}{ALPHA}), \bibinfo{journal}{Nucl. Phys. B}
  \textbf{\bibinfo{volume}{545}}, \bibinfo{pages}{529} (\bibinfo{year}{1999}),
  \eprint{hep-lat/9808013}.

\bibitem[{\citenamefont{Caswell}(1974)}]{Caswell:1974gg}
\bibinfo{author}{\bibfnamefont{W.~E.} \bibnamefont{Caswell}},
  \bibinfo{journal}{Phys. Rev. Lett.} \textbf{\bibinfo{volume}{33}},
  \bibinfo{pages}{244} (\bibinfo{year}{1974}).

\bibitem[{\citenamefont{Jones}(1974)}]{Jones:1974mm}
\bibinfo{author}{\bibfnamefont{D.~R.~T.} \bibnamefont{Jones}},
  \bibinfo{journal}{Nucl. Phys. B} \textbf{\bibinfo{volume}{75}},
  \bibinfo{pages}{531} (\bibinfo{year}{1974}).

\bibitem[{\citenamefont{Bode et~al.}(2000)\citenamefont{Bode, Weisz, and
  Wolff}}]{Bode:1999sm}
\bibinfo{author}{\bibfnamefont{A.}~\bibnamefont{Bode}},
  \bibinfo{author}{\bibfnamefont{P.}~\bibnamefont{Weisz}}, \bibnamefont{and}
  \bibinfo{author}{\bibfnamefont{U.}~\bibnamefont{Wolff}}
  (\bibinfo{collaboration}{ALPHA}), \bibinfo{journal}{Nucl. Phys. B}
  \textbf{\bibinfo{volume}{576}}, \bibinfo{pages}{517} (\bibinfo{year}{2000}),
  \bibinfo{note}{[Erratum: Nucl.Phys.B 608, 481--481 (2001), Erratum:
  Nucl.Phys.B 600, 453--453 (2001)]}, \eprint{hep-lat/9911018}.

\bibitem[{\citenamefont{Plasencia
  et~al.}(2021{\natexlab{b}})\citenamefont{Plasencia, Brida, de~Divitiis,
  Lytle, Papinutto, Pirelli, and Vladikas}}]{Plasencia:2021ihd}
\bibinfo{author}{\bibfnamefont{I.~C.} \bibnamefont{Plasencia}},
  \bibinfo{author}{\bibfnamefont{M.~D.} \bibnamefont{Brida}},
  \bibinfo{author}{\bibfnamefont{G.~M.} \bibnamefont{de~Divitiis}},
  \bibinfo{author}{\bibfnamefont{A.}~\bibnamefont{Lytle}},
  \bibinfo{author}{\bibfnamefont{M.}~\bibnamefont{Papinutto}},
  \bibinfo{author}{\bibfnamefont{L.}~\bibnamefont{Pirelli}}, \bibnamefont{and}
  \bibinfo{author}{\bibfnamefont{A.}~\bibnamefont{Vladikas}}, in
  \emph{\bibinfo{booktitle}{{38th International Symposium on Lattice Field
  Theory}}} (\bibinfo{year}{2021}{\natexlab{b}}), \eprint{2111.15325}.

\end{thebibliography}
\end{document}